\documentclass[11pt]{article} 
\usepackage[top=20mm,bottom=20mm, left=20mm, right=20mm]{geometry}
\usepackage{graphicx,amsmath,amssymb,url,enumerate,mathrsfs,epsfig,color,subfigure,transparent,import,hyperref, cite}
\hypersetup{
    colorlinks=true,       
    linkcolor=red,          
    citecolor=blue,        
    filecolor=magenta,      
    urlcolor=cyan           
}

\usepackage{chngcntr}
 
\newtheorem{prop}{Proposition}[section]
\newtheorem{thm}{Theorem}[section]

\newtheorem{rem}{\it Remark}[section]

\title{Non-metric connection and metric anomalies in materially uniform elastic solids}
\author{Ayan Roychowdhury and Anurag Gupta\thanks{ag@iitk.ac.in}}
\date{{\small Department of Mechanical Engineering, 
Indian Institute of Technology, Kanpur 208016, 
India.\\ \today}}
\begin{document}
\maketitle

\begin{abstract}
Metric anomalies arising from a distribution of point defects (intrinsic interstitials, vacancies, point stacking faults), thermal deformation, biological growth, etc. are well known sources of material inhomogeneity and internal stress. By emphasizing the geometric nature of such anomalies we seek their representations for materially uniform crystalline elastic solids. In particular, we introduce a quasi-plastic deformation framework where  the multiplicative decomposition of the total deformation gradient into an elastic and a plastic deformation is established such that the plastic deformation is further  decomposed  multiplicatively in terms of a deformation due to dislocations and another due to metric anomalies. We discuss our work in the context of quasi-plastic strain formulation and Weyl geometry. We also derive a general form of metric anomalies which yield a zero stress field in the absence of other inhomogeneities and any external sources of stress. 
\end{abstract}




\section{Introduction}

The purpose of this article is to discuss several issues regarding the geometric nature of metric anomalies in materially uniform simple elastic solids. Metric anomalies appear in a non-Riemannian geometric space whenever the inner product of any two tangent vectors is not preserved under parallel transport. They are represented by a non-trivial non-metricity tensor field in the geometric space \cite{schouten}. In the context of elastic solids, where the relevant geometric space is the stress-free material space, they can be identified with material inhomogeneity fields arising from a distribution of point defects (intrinsic interstitials, vacancies, point stacking faults), thermal deformation, biological growth, etc. \cite{kroner81a, kroner94, bilby66,  anth71, miri,yavari12, yavari14, yavari2014a}. This association is made on the basis of the metrical nature of these material inhomogeneities, such as that leading to an inhomogeneous volume change due to a distribution of spherical point defects, isotropic thermal deformation or isotropic growth. The correspondence of non-metricity with metric anomalies is analogous to that of torsion of the material space with dislocation density \cite{kondo52, bilby55} and curvature of the material space with disclination density \cite{anth70}. Both dislocations and metric anomalies are commonly observed sources of material inhomogeneity in crystalline elastic solids. A precise understanding of the geometric nature of the inhomogeneity distribution is essential for posing meaningful boundary-value-problems to determine residual stresses and shape changes in materially inhomogeneous solid bodies \cite{kroner81a}. 

\begin{figure}[t!]
 \centering
 \includegraphics[scale=0.45]{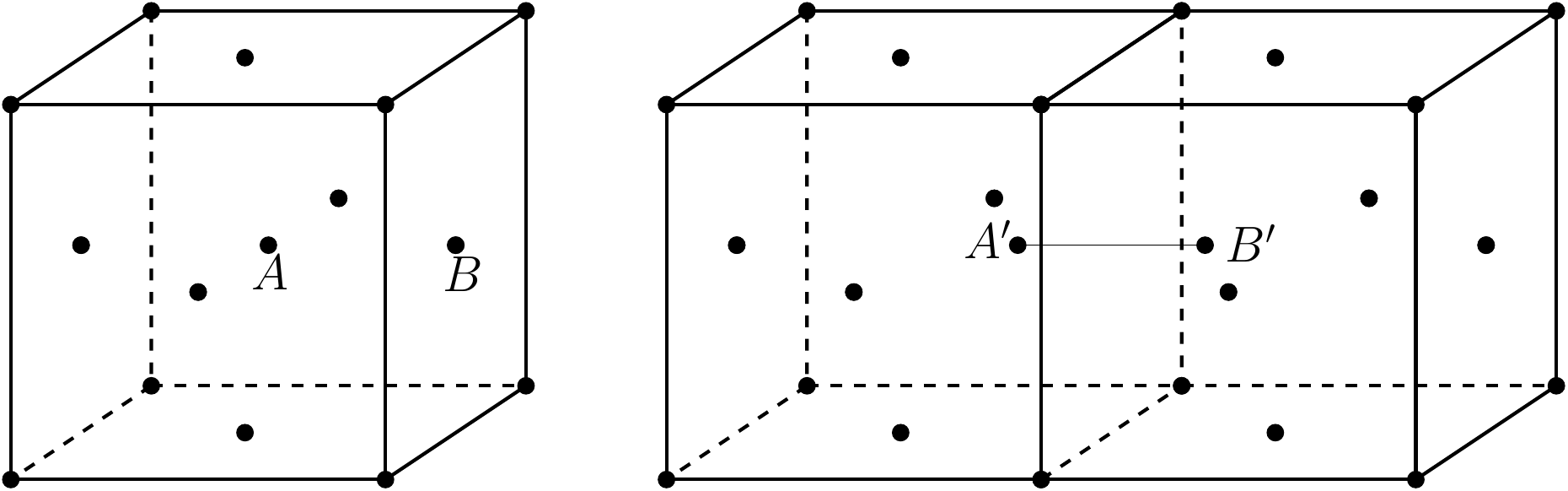}
\caption{Schematic diagram of a split-interstitial in fcc lattice. The original interstitial, located at $A$, is unstable and relaxes into a dumbbell shaped split-interstitial $A'B'$ (reproduced from \cite{esh4}).}
 \label{split}
\end{figure}

\begin{figure}[t!]
 \centering
 \subfigure[]{\includegraphics[scale=0.45]{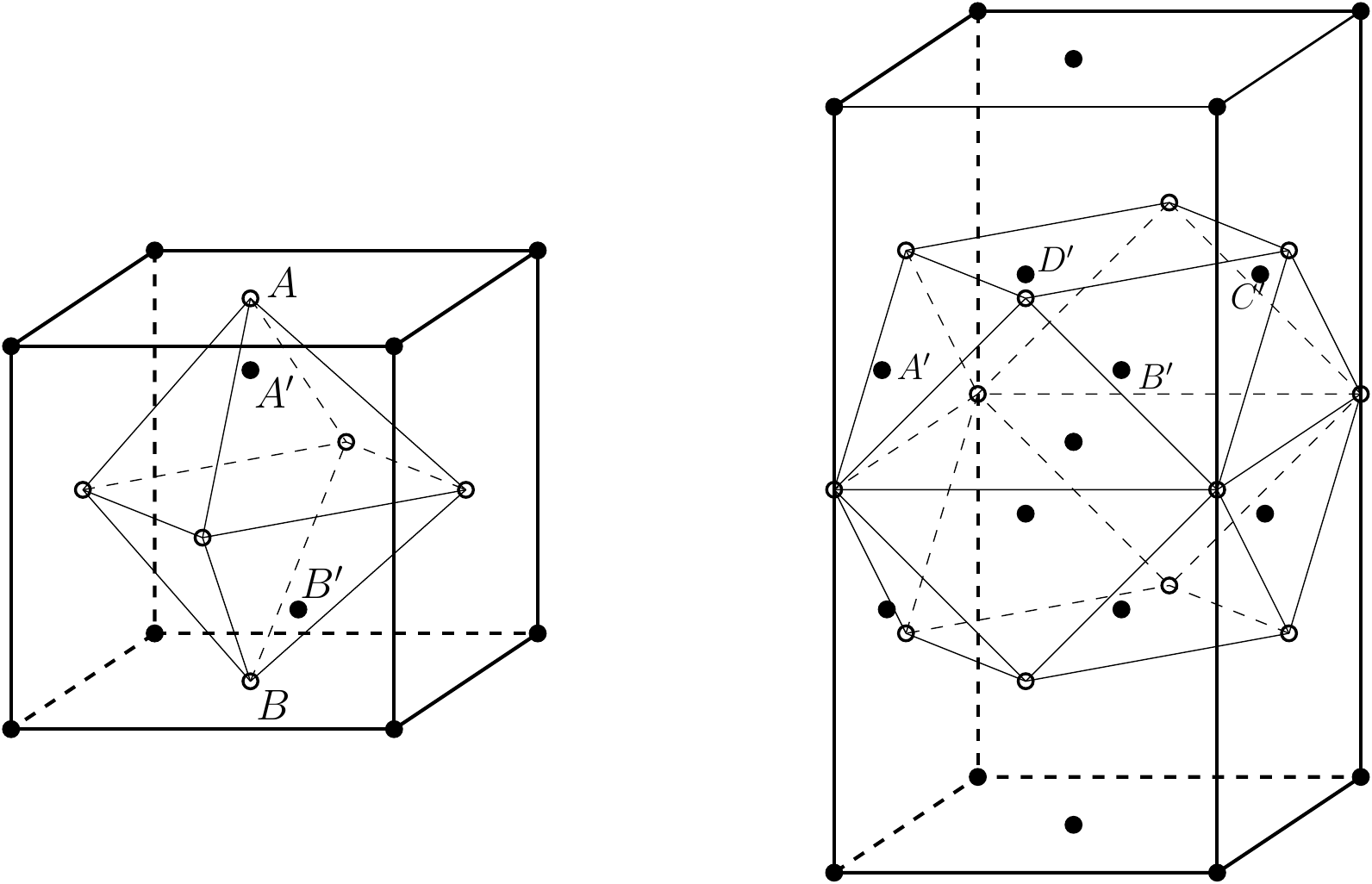}}
 \hspace{10mm}
 \subfigure[]{\includegraphics[scale=0.45]{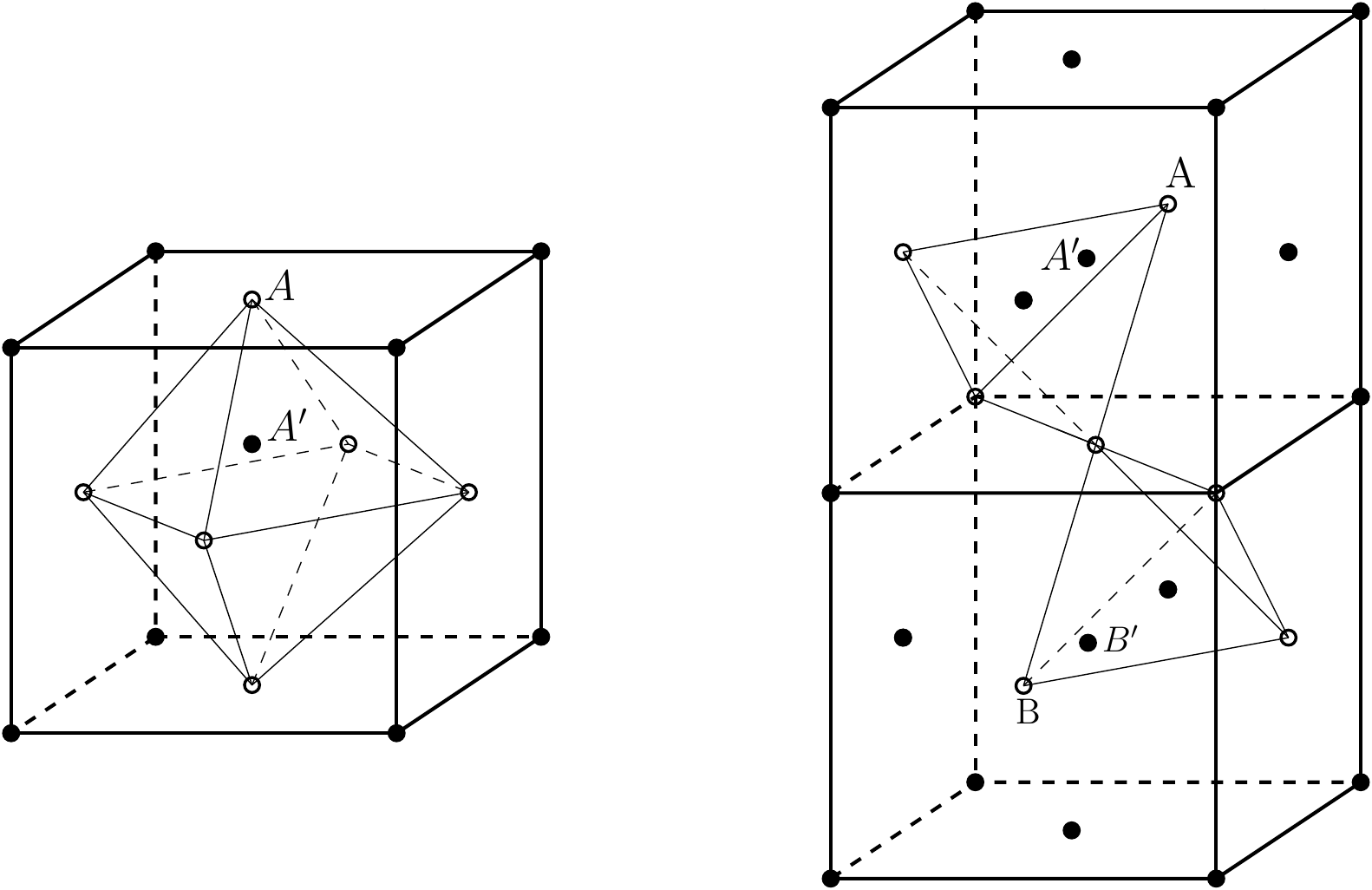}}
 \caption{Schematic diagrams of a (a) tetra and a (b) penta vacancy in Copper in their stable configurations (reproduced from \cite{vineyard}).}
\label{tetra-penta}
 \end{figure}

We are in particular interested in representations of anisotropic metric anomalies. This is in contrast to the more popular isotropic descriptions of a distribution of spherically symmetric point defects \cite{miri,yavari12, yavari14, yavari2014a} and isotropic thermal expansion coefficient. It is well known that stable configurations of clusters of point defects form exotic anisotropic shapes \cite{kiritani99 ,kiritani2000, vineyard}. In these works, divacancies have been found to be more mobile than single vacancies and clusters of trivacancies in Copper stronger, with increased  binding energy, against separation into single vacancies. As reported by Kiritani et al. \cite{kiritani99, kiritani2000}, high density of small vacancy clusters in the form of stacking-fault tetrahedra dominate the plastic deformation of thin
foils of fcc materials under high strain-rate without any intervention from dislocations. Anisotropic cluster formation is usually more stable than free standing spherically symmetric defects.
 For example, as shown in Figure \ref{split}, an intrinsic interstitial atom in fcc lattice relaxes into a split-interstitial in order to achieve stability. The interstitial atom at the original position $A$ pushes the atom at $B$ towards right yielding a dumbbell shaped defect $A'B'$ in the stable state. Rather than modelling point defects as spherically symmetric objects, it is only appropriate to consider point defect density as a distribution of infinitesimal rod-like dumbbell structures in the crystalline body. The individual dumbbells display transverse isotropic symmetry about their axes. In Figure \ref{tetra-penta}, stable configurations of tetra and penta-vacancies in fcc Copper are depicted. The stable configurations are of octahedral and decaoctahedral shape for tetra vacancies, and of octahedral and bi-tetrahedral shape for penta vacancies. A continuous description of these  clusters of vacancies requires an anisotropic representation of metric anomalies. Another example of elementary anisotropic point defects was described by Kr\"oner \cite{kroner90, kroner94} in the form of point stacking faults. In this case, the individual point stacking faults are themselves anisotropic. Further instances of anisotropic metric anomaly are provided by finite thermal deformation in crystalline materials \cite{barron} and bulk growth in biological materials \cite{yavari2010}. In these cases, the thermal expansion or the growth coefficient is described by an anisotropic, symmetric second order tensor. It is prudent to emphasize here that the anisotropy we are referring to is the anisotropy associated with the structural symmetry of the distributed inhomogeneity (metric anomalies in the present case), which is independent of the symmetry (anisotropic or otherwise) of the material response function. For instance, a materially uniform body with isotropic material response can contain a distribution of anisotropic point defects such as those shown in Figures \ref{split} and \ref{tetra-penta}.

In the first part of this article (Section \ref{sec2}) we introduce the notion of material space by attributing a metric and a non-metric affine connection, with non-zero torsion and curvature, to the 3-dimensional body manifold of a materially uniform simple elastic solid. The metric and the connection are both constructed from a given distribution of inhomogeneities in the body and an assumed constitutive response. In doing so, we extend the formulation of Noll \cite{noll} and Wang \cite{wang}, which is restricted to a dislocated material body, where the material connection and metric are derived solely from the constitutively determined material uniformity field. We describe the geometrical significance of torsion, curvature and non-metricity, and relate them to distributions of dislocations, disclinations and metric anomalies, respectively. The main result in this section is the development of the notion of {\it metrical disclinations} and their relation with metric anomalies (see Proposition \ref{piq}). Metrical disclinations can appear only in a non-metric space and are related to path dependence of the inner product of tangent vectors. Unlike the well known rotational disclinations, which are the only kind of disclinations possible in metric-compatible spaces, metrical disclinations are not fundamental line defects in materially uniform simple elastic solids. A distribution of rotational disclinations is also unfeasible in crystalline solids due to their unrealistically high elastic energy. Motivated by these concerns, we look for simplified representations of non-metricity in the absence of curvature in material space. This is equivalent to requiring distant material parallelism for crystalline solids.

In the second part of the paper (Section \ref{sec3}) we focus on obtaining rigorous results of representations for non-metricity tensor for a zero curvature space. Towards this end, we use the third Bianchi-Padova relation to obtain a necessary and sufficient representation of non-metricity in terms of a symmetric second order tensor (see Proposition \ref{uni}). This also leads us to introduce the auxiliary material space which inherits the affine connection from the material space but has a metric such that the non-metricity vanishes identically. In particular, we recover the quasi-plastic strain framework, proposed by Anthony \cite{anth71}, now established on firm geometrical grounds. We also discuss non-metricity in the context of semi-metric geometry (which with zero torsion is called Weyl geometry). We show that the non-metricity tensor therein necessarily has an isotropic form, given in terms of a scalar field, when curvature of the space is identically zero (see Proposition \ref{semimetric}).     As a result, the Weyl geometry framework in its standard form, where the Weyl co-vector form is exact leading to an isotropic form of non-metricity \cite{yavari12, yavari14}, is insufficient to model anisotropic metric anomalies.

We propose a novel representation of metric anomalies in terms of a second order tensor (we call it {\it quasi-plastic deformation}) such that the total deformation gradient (with respect to a fixed reference configuration) can be multiplicatively decomposed into an elastic and a plastic deformation. Moreover, the plastic deformation is further decomposed multiplicatively in terms of a dislocation induced deformation and the quasi-plastic deformation (see Proposition \ref{qpdprop}). Such a framework is amenable to analytical and numerical solutions of boundary-value-problems for (internal) stress and displacement fields for elastic solids having a continuous distribution of dislocations and metric anomalies. The representation of non-metricity in terms of quasi-plastic deformation also allows us to consider a broader range of metric anomalies than what is afforded by quasi-plastic strain framework. 

Finally, before concluding in Section \ref{sec5}, we digress briefly in Section \ref{sec4} to obtain the general form of non-metricity tensor which corresponds to a zero stress field in the absence of dislocations, disclinations and any external source of stress. We derive a closed form solution to this problem in a linearized situation, assuming non-metricity and elastic strain to be small and of the same order (see Proposition \ref{nilnm}).

\section{Material response function and associated geometric constructions}\label{sec2}
Our prototype for the theory of a continuous material body is a 3-dimensional differential manifold $\mathcal{B}$ which can be covered with a single chart. $\mathcal{B}$ is classically known as the \textit{material manifold} \cite{noll, wang} and the points in $\mathcal{B}$, designated by $\boldsymbol{X}$, are called \textit{material points}. We assume the manifold structure on $\mathcal{B}$ to be sufficiently differentiable as the context demands and use a holonomic curvilinear coordinate system $\theta^i$ to label the material points $\boldsymbol{X}\in\mathcal{B}$. Roman indices ($i$, $j$, $p$ etc.) take the values 1, 2 and 3, and Einstein's summation convention holds over repeated indices. From its manifold structure, $\mathcal{B}$ naturally inherits the Euclidean properties of $\mathbb{R}^3$, including the Euclidean inner product (denoted by $\boldsymbol{\cdot}$). Let $\boldsymbol{G}_i$ be the natural basis vector field of the coordinate system $\theta^i$, $G_{ij}:=\boldsymbol{G}_i\boldsymbol{\cdot}\boldsymbol{G}_j$ be the components of the Euclidean metric tensor with respect to the coordinates $\theta^i$, $[G^{ij}]:=[G_{ij}]^{-1}$ and $\boldsymbol{G}^i:= G^{ij}\boldsymbol{G}_j$ the natural co-vector basis field.

With the tangent space $T_{\boldsymbol{X}}\mathcal{B}$ of $\mathcal{B}$ at $\boldsymbol{X}$ as the underlying vector space, we denote $Lin$ and $InvLin^+$ as the set of all  second order tensors and invertible second order tensors with positive determinant, respectively, $Sym$ and $Sym^+$ as the set of all second order symmetric and symmetric positive definite tensors, respectively, $Skw$ as the set of all second order skew symmetric tensors, $Unim$ as the set of all second order tensors with determinant equal to 1 and $Orth^+$ as the set of all proper orthogonal second order tensors (i.e., rotations). We denote the identity tensor field over the manifold $\mathcal{B}$ by $\boldsymbol{I}:=G_{ij}\boldsymbol{G}^i\otimes\boldsymbol{G}^j=G^{ij}\boldsymbol{G}_i\otimes\boldsymbol{G}_j$. The inverse of an invertible tensor is indicated by a superscript $(-1)$ while the transpose is denoted by a superscript $T$. We use the shorthand notation $(\cdot)_{,i}$ for the ordinary partial derivative $\frac{\partial(\cdot)}{\partial \theta^i}$.

\subsection{Material uniformity} \label{uniform}

We restrict our consideration to materials classically known as \textit{simple elastic solids} (without heat conduction). The constitutive response function for such a material is given by a mapping
\begin{equation}
\hat W: Sym^+\times\mathcal{B}\to\mathbb{R}^+,
\end{equation} 
known as the strain energy density function. Here, $\mathbb{R}^+$ denotes the set of non-negative real numbers. In addition, the body is assumed to be {\it materially uniform} in the sense that for every pair $\boldsymbol{X},\boldsymbol{Y}\in\mathcal{B}$, there exists a second order tensor $\boldsymbol{K}_{\boldsymbol{XY}}:T_{\boldsymbol{X}}\mathcal{B}\to T_{\boldsymbol{Y}}\mathcal{B}$, with $\mbox{det}\,\boldsymbol{K}_{\boldsymbol{XY}}>0$ (det denotes the determinant operator), such that 
\begin{equation}
\hat W(\boldsymbol{K}^T_{\boldsymbol{XY}}\,\boldsymbol{h}\,\boldsymbol{K}_{\boldsymbol{XY}},\boldsymbol{X})=\hat W(\boldsymbol{h},\boldsymbol{Y})
\label{uniformity}
\end{equation}
is satisfied for all $\boldsymbol{h}\in Sym^+$.\footnote{The domain of the partial function $\hat W(\cdot,\boldsymbol{X})$, for $\boldsymbol{X}\in \mathcal{B}$, is customarily assumed to be $InvLin^+$, the space where the \textit{deformation gradients} reside \cite{noll,wang}, which, under the Principle of Material Frame Indifference, gets restricted to its subset $Sym^+$. In presence of certain material inhomogeneities (e.g., disclinations), a well-defined element in $InvLin^+$ may not exist to appear in the constitutive function. Our treatment bypasses this limitation, as it is always guaranteed that a well-defined element in $Sym^+$ exists to appear in $\hat W(\cdot,\boldsymbol{X})$ as an argument. This well-defined element could be, in our context, any of the standard measures of strain.} It can be shown that the set of values of $\boldsymbol{K}_{\boldsymbol{XY}}$ satisfying \eqref{uniformity}, for fixed $\boldsymbol{X,Y}\in\mathcal{B}$, forms a group $\mathcal{K}_{\boldsymbol{XY}}$. Moreover, the material symmetry group at $\boldsymbol{X}$, defined as $\mathcal{G}_{\boldsymbol{X}}:=\mathcal{K}_{\boldsymbol{XX}}$, in order to conform to the mass consistency condition, must satisfy $\mathcal{G}_{\boldsymbol{X}}\subseteq Unim$ \cite{noll}.

Fixing a material point $\boldsymbol{X}_0\in\mathcal{B}$ in the materially uniform body, we can define a field $\boldsymbol{K}(\boldsymbol{X}):=\boldsymbol{K}_{\boldsymbol{X}_0 \boldsymbol{X}}$ that satisfies
\begin{equation}
W_{\boldsymbol{X}_0}(\boldsymbol{K}^T(\boldsymbol{X})\,\boldsymbol{h}\,\boldsymbol{K}(\boldsymbol{X}))=\hat W(\boldsymbol{h},\boldsymbol{X})
\label{uniformity1}
\end{equation}
for all $\boldsymbol{h}\in Sym^+$, where $W_{\boldsymbol{X}_0}:Sym^+\to\mathbb{R}^+$ is defined as $W_{\boldsymbol{X}_0}(\cdot):=\hat W (\cdot,\boldsymbol{X}_0)$. $\boldsymbol{K}$ is known as the material uniformity field with respect to the material point $\boldsymbol{X}_0$ \cite{epsbook}. Since the body is assumed to be materially uniform, the choice of the material point $\boldsymbol{X}_0$ is arbitrary. This renders the constitutive response function independent of any explicit dependence on material points, as is clear from the expression \eqref{uniformity1}. The body is called a materially uniform solid, if we can choose such a $\boldsymbol{X}_0$ such that $\mathcal{G}_{\boldsymbol{X}_0}\subseteq Orth^+$.\footnote{Apart from the point symmetry group $\mathcal{G}$, which essentially describes rotational symmetries, the material structure presently under consideration possesses, due to its expanse in the Euclidean 3-space, spatial translational symmetries.} In the present work, we restrict ourselves to materially uniform simple elastic solids.

\subsection{Material $\mathcal{G}$-structure, material connection and material metric} \label{gstructure}
The material uniformity field $\boldsymbol{K}$ appearing in \eqref{uniformity1} is, in general, multi-valued due to non-trivial symmetry groups at $\boldsymbol{X}$ as well as at $\boldsymbol{X}_0$. A fibre bundle can be constructed by attaching the values of $\boldsymbol{K}(\boldsymbol{X})$ at respective $\boldsymbol{X}\in\mathcal{B}$, giving rise to the {\it material $\mathcal{G}$-structure} \cite{wang,epsbook}. It can be shown that the material $\mathcal{G}$-structure is a principal fibre bundle, with structure group (which is the same as the standard fibre) $\mathcal{G}:=\mathcal{G}_{\boldsymbol{X}_0}$. The domain of $\boldsymbol{K}$, given a fixed degree of differentiability $C^k$, may not span the whole material manifold $\mathcal{B}$. The material $\mathcal{G}$-structure is hence, in general, non-trivial.

For geometric constructions on the material $\mathcal{G}$-structure, it can be equipped with an arbitrary affine connection and metric. However, in order to formulate a geometric theory of the underlying material structure, we choose only a particular affine connection $\mathfrak{L}$ and a particular metric $\boldsymbol{g}$ out of these infinite possibilities, as informed by the inhomogeneities present in the material structure, if any. The fundamental geometric objects associated with $\mathfrak{L}$ and $\boldsymbol{g}$, namely, the torsion tensor $\mathfrak{T}$, the curvature tensor $\mathfrak{R}$ and the non-metricity tensor $\mathfrak{Q}$ can be naturally identified with the densities of dislocations, disclinations and metric anomalies, respectively, as we will see in the following. Once these identifications are made, and the defect densities in a given material body are known in terms of these fundamental geometric objects over $\mathcal{B}$, the connection $\mathfrak{L}$ and the metric $\boldsymbol{g}$ can be constructed from these geometric objects by solving a system of PDEs. This system is constituted of the respective defining equations of the tensors $\mathfrak{T}$, $\mathfrak{R}$ and $\mathfrak{Q}$ in terms of $\mathfrak{L}$ and $\boldsymbol{g}$.

For instance, the governing equation for $\mathfrak{L}$, for an appropriately specified tensor $\mathfrak{R}$ (the density of disclinations), is a system of first order non-linear PDEs. From a result by Talvacchia \cite[Theorem 7]{jtalva} on the existence of an affine connection over a $Unim$-principal bundle\footnote{A $Unim$-principal bundle is a principal fibre bundle whose structure group is the group $Unim$.} with 3-dimensional base manifold whose curvature tensor is prescribed {\it a priori} as a generic real analytic function, and from the fact that the material $\mathcal{G}$-structure is a principal subbundle of a $Unim$-principal bundle (since $\mathcal{G}\subseteq Orth^+ \subset Unim$) with the 3-dimensional base manifold $\mathcal{B}$, it follows that a solution $\mathfrak{L}$ exists, provided that $\mathfrak{R}$ is analytic which we assume to be the case here. The torsion tensor obtained from this affine connection $\mathfrak{L}$ has to be equal to an appropriately specified tensor $\mathfrak{T}$ (the density of dislocations). On the other hand, the governing equation for $\boldsymbol{g}$ is a system of first order non-homogeneous linear PDEs, given appropriately the functions $\mathfrak{Q}$ (the density of metric anomalies) and $\mathfrak{L}$. If we assume $\mathfrak{Q}$ and the previously obtained $\mathfrak{L}$, as just discussed, to be analytic, then a unique analytic solution $\boldsymbol{g}$ indeed exists as a consequence of the Cauchy-Kowalevski and Holmgren's existence and uniqueness theorems for first order linear system of PDEs with analytic coefficients and data.\footnote{Within the realm of the classical solutions of the PDEs that we are considering here, the existence and uniqueness theorems of Cauchy-Kowalevski and Holmgren do no extend to the class of smooth functions which are not analytic; in this context, we would like to refer to the well-known Lewy's example that demonstrates a linear PDE with smooth coefficients which has no solution \cite{lewy}.} We will construct an explicit solution for $\mathfrak{L}$ and $\boldsymbol{g}$ involving the material uniformity field $\boldsymbol{K}$ and other physically relevant quantities in Section \ref{secqpd}.

The affine connection $\mathfrak{L}$, thus constructed, is called the {\it material connection}, and the metric $\boldsymbol{g}$, the {\it material metric}. The ``material'' nature of this connection and metric is clear from the above discussion; the fundamental geometric objects they yield, viz. the torsion, curvature and non-metricity, represent various inhomogeneity measures present in the material structure of the body.

\begin{rem}
 {\rm
The above construction of material connection and material metric generalizes the original idea of Noll \cite{noll} and Wang \cite{wang}, which is purely constitutive in nature and is applicable to dislocated material bodies free from any curvature and metric anomalies. The general form of a zero-curvature connection is determined solely in terms of an invertible second order tensor field which, in turn, determines the metric compatible with the connection (see Section \ref{secqpd} for details); Noll identified this second order tensor with the material uniformity field $\boldsymbol{K}$. In the presence of curvature and metric anomalies, material connection and metric can no longer be derived from the material uniformity field alone but require additional information which comes from a given distribution of curvature and metric anomalies. In fact, with whatever connection and metric one is adorning the material $\mathcal{G}$-structure, they should be compatible with the underlying constitutive nature of the body and the contained inhomogeneities.
 }
\end{rem}

\subsubsection*{The material space}
Manifold $\mathcal{B}$, equipped with the material connection $\mathfrak{L}$ and the material metric $\boldsymbol{g}$, forms the \textit{material space}. We denote the material space by the triple $(\mathcal{B},\mathfrak{L},\boldsymbol{g})$ and the coefficients of $\mathfrak{L}$ and the components of $\boldsymbol{g}$, with respect to the embedded coordinate system $\theta^i$, by $L^i_{kj}$ and $g_{ij}$, respectively.  The raising and lowering of indices of components of tensorial objects are performed with respect to the purely covariant components $g_{ij}$ and the purely contravariant components $g^{ij}$, where $[g^{ij}]:=[g_{ij}]^{-1}$. The covariant differentiation of a quantity with respect to $\mathfrak{L}$ is denoted by $\nabla$, for example,
\begin{equation}
 \nabla_{j}u^i:=u^i_{,j}+L^i_{jk}u^k,\,\,\,\,\nabla_{j}u_i:=u_{i,j}-L^k_{ji}u_k \,\,\,\textrm{etc.},
\end{equation}
where all the components are with respect to the coordinates $\theta^i$.

Given a tangent vector with components $u^i_0$ at the initial point of a curve $\mathcal{C}:=\{C^i(s)\boldsymbol{G}_i (s) \in \mathcal{B}\,\big|\,s\in[0,1]\}$, a {\it materially constant} tangent vector field $u^i(C^k(\tau))$ is constructed by solving the linear ODE
\begin{equation}
 \frac{du^i(\tau)}{d\tau}=-L^i_{kj}(\tau)u^j(\tau) \dot{C}^k(\tau),\,\,\,\,\,u^i(\tau=0)=u^i_0.
\end{equation}
In other words, a materially constant vector field on a curve, by definition, has zero directional covariant derivative with respect to the material connection throughout the curve.

It is evident from our construction of the material connection $\mathfrak{L}$ and the material metric $\boldsymbol{g}$, as discussed previously, that the fundamental geometric objects of the material space $(\mathcal{B},\mathfrak{L},\boldsymbol{g})$ provide natural measures for various material inhomogeneities contained within the body. Importantly, the material inhomogeneity densities remain unaffected by a superposed compatible deformation of the body. In other words, material space is a geometric space where an internal observer would be able to detect only configurational changes in the body (i.e. those arising out of defects) but would otherwise fail to observe any compatible deformations incurred by the body as a result of external loading, etc. \cite{kroner81a}. 


\subsection{Material torsion tensor}
\begin{figure}[t!]
 \centering
 \subfigure[]{\includegraphics[scale=0.55]{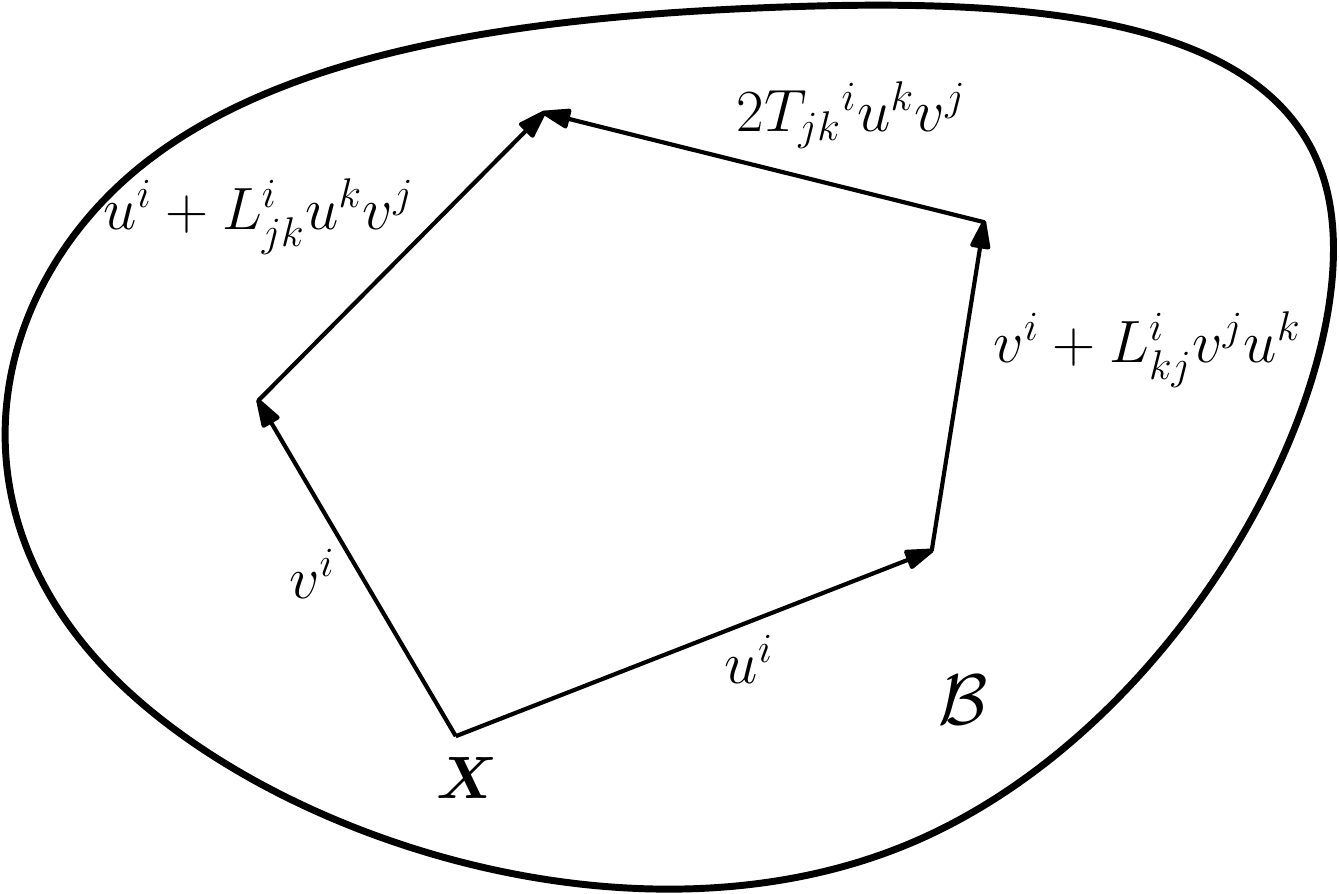}}
 \hspace{10mm}
 \subfigure[]{\includegraphics[scale=0.55]{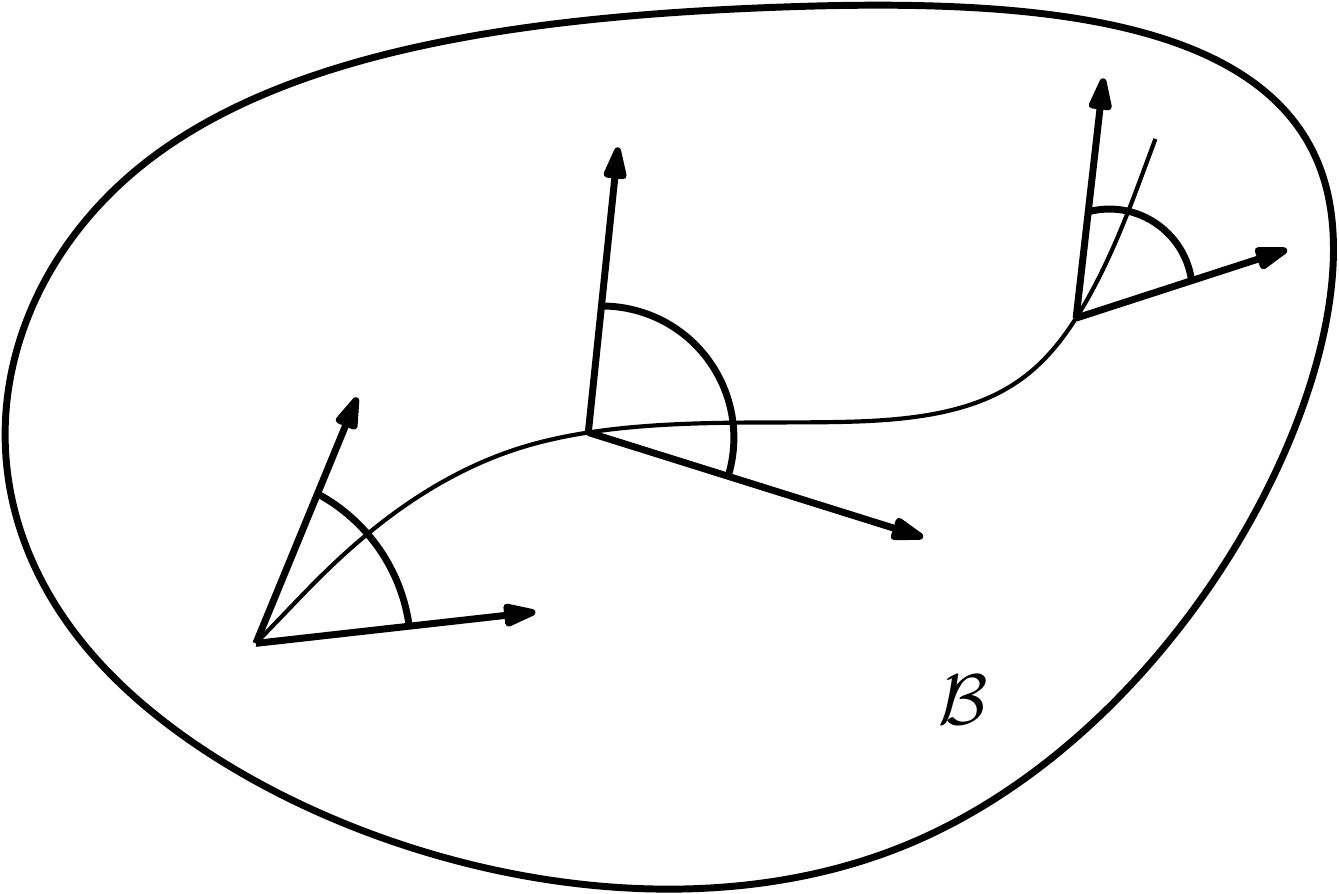}}
 \caption{(a) Closure failure of an infinitesimal parallelogram due to torsion. (b) Change in length and angle between two vectors under parallel transport due to non-metricity.}
 \label{torsion&non-metricity}
\end{figure}

The third order torsion tensor $\mathfrak{T}$ associated with the material connection $\mathfrak{L}$ is a mapping
\begin{equation}
 \mathfrak{T}:T_{\boldsymbol{X}}\mathcal{B}\times T_{\boldsymbol{X}}\mathcal{B}\to T_{\boldsymbol{X}}\mathcal{B},
\end{equation}
which is bilinear and skew with respect to its arguments. Its components $T_{jk}{}^i$ with respect to the coordinates $\theta^i$ are defined as
\begin{equation}
 T_{jk}{}^i:=L^i_{[jk]}.
 \label{def:torsion}
\end{equation}
Here, the square bracket in the subscript indicates the skew part of the field with respect to the enclosed indices (whereas a round bracket is used to indicate the symmetric part). Torsion tensor measures the closure failure of an infinitesimal parallelogram in the material manifold, and it is one of the fundamental geometric objects on the material space. The construction of such a parallelogram is illustrated in Figure~\ref{torsion&non-metricity}(a).

\subsubsection*{Torsion inhomogeneities} Torsion tensor of the material manifold is the natural measure for density of dislocations in the material body (first identified by Kondo~\cite{kondo52} and, later, independently by  Bilby et al.~\cite{bilby55}). Dislocations are one of the fundamental line defects in materially uniform simple elastic solids; they are associated with the translational symmetries of the underlying material structure. This identification is evident from the similar nature of the two objects, viz.~the closure failure of infinitesimal parallelograms in the material space and the closure failure of the Burgers circuit \cite{kroner81a}. The second order axial tensor of torsion, which has components $\alpha^{ij}:=\varepsilon^{imn}T_{mn}{}^j$, is called the dislocation density tensor. The diagonal components of the matrix $[\alpha^{ij}]$ measure the density of edge dislocations and the off-diagonal components measure the density of screw dislocations \cite{kroner81a}.

\subsection{Material curvature tensor}\label{matcurv}

The fourth order Riemann-Christoffel curvature tensor $\mathfrak{R}$ of the material connection $\mathfrak{L}$ is a mapping
\begin{equation}
\mathfrak{R}:T_{\boldsymbol{X}}\mathcal{B}\times T_{\boldsymbol{X}}\mathcal{B}\to Lin,
\end{equation}
which is bilinear and skew with respect to its arguments. Its components $R_{jiq}{}^p$ with respect to the coordinates $\theta^i$ are defined as
\begin{equation}
 R_{jiq}{}^p := L^p_{iq,j}- L^p_{jq,i}+L^h_{iq} L^p_{jh}-L^h_{jq} L^p_{ih}.
 \label{def:curvature}
\end{equation}
The curvature tensor $\mathfrak{R}$ measures, in the linear approximation, the change that a tangent vector suffers under parallel transport along an infinitesimal loop. It is our second fundamental geometric object on the material space.

We also define the purely covariant components $R_{klji}$ of $\mathfrak{R}$, by lowering the fourth index with the material metric $g_{ij}$, as
\begin{equation}
 R_{klji}:=g_{ip} R_{klj}{}^p.
 \label{def:covcurvature}
\end{equation}
It follows from the definition that $R_{klq}{}^p=-R_{lkq}{}^p$ and $R_{klij}=-R_{lkij}$.

\subsubsection*{Decomposition of the curvature tensor}
it is useful for our present objective to decompose the components $R_{klij}$ as (cf. \cite{povs})
 \begin{equation}
  R_{klij}=\varepsilon_{pkl}\varepsilon_{qij}\theta^{pq}+\varepsilon_{pkl}\zeta_{ij}{}^p,
  \label{curv:decomposition}
 \end{equation}
where $\theta^{pq}$ and $\zeta_{ij}{}^p$ are defined as
\begin{equation}
\theta^{pq}:=\frac{1}{4}\varepsilon^{pij}\varepsilon^{qkl}R_{ij[kl]}\,\,\bigg(=\frac{1}{4}\varepsilon^{pij}\varepsilon^{qkl}R_{ijkl}\bigg)\,\,\,\,\,\textrm{and}\,\,\,\,\,\zeta_{ij}{}^p:=\frac{1}{2}\varepsilon^{pkl}R_{kl(ij)}.
\end{equation}
Here, $\varepsilon^{ijk}:=g^{-\frac{1}{2}}e^{ijk}$ and $\varepsilon_{ijk}:=g^{\frac{1}{2}}e_{ijk}$, where $e^{ijk}=e_{ijk}$ is the 3-dimensional permutation symbol and $g:=\mbox{det}[g_{ij}]$. The second order tensor field $\boldsymbol{\theta}:=\theta^{pq}\boldsymbol{G}_p\otimes\boldsymbol{G}_q$ characterizes the skew part and the third order tensor field $\boldsymbol{\zeta}:=\zeta_{ij}{}^p\boldsymbol{G}_p\otimes\boldsymbol{G}^i\otimes\boldsymbol{G}^j$ characterizes the symmetric part of the tensor field $\mathfrak{R}(\cdot,\cdot)\in {Lin}$. The geometric interpretation of the symmetric and skew part is illustrated in the following.

\begin{figure}[t!]
 \centering
 \subfigure[]{\includegraphics[scale=0.55]{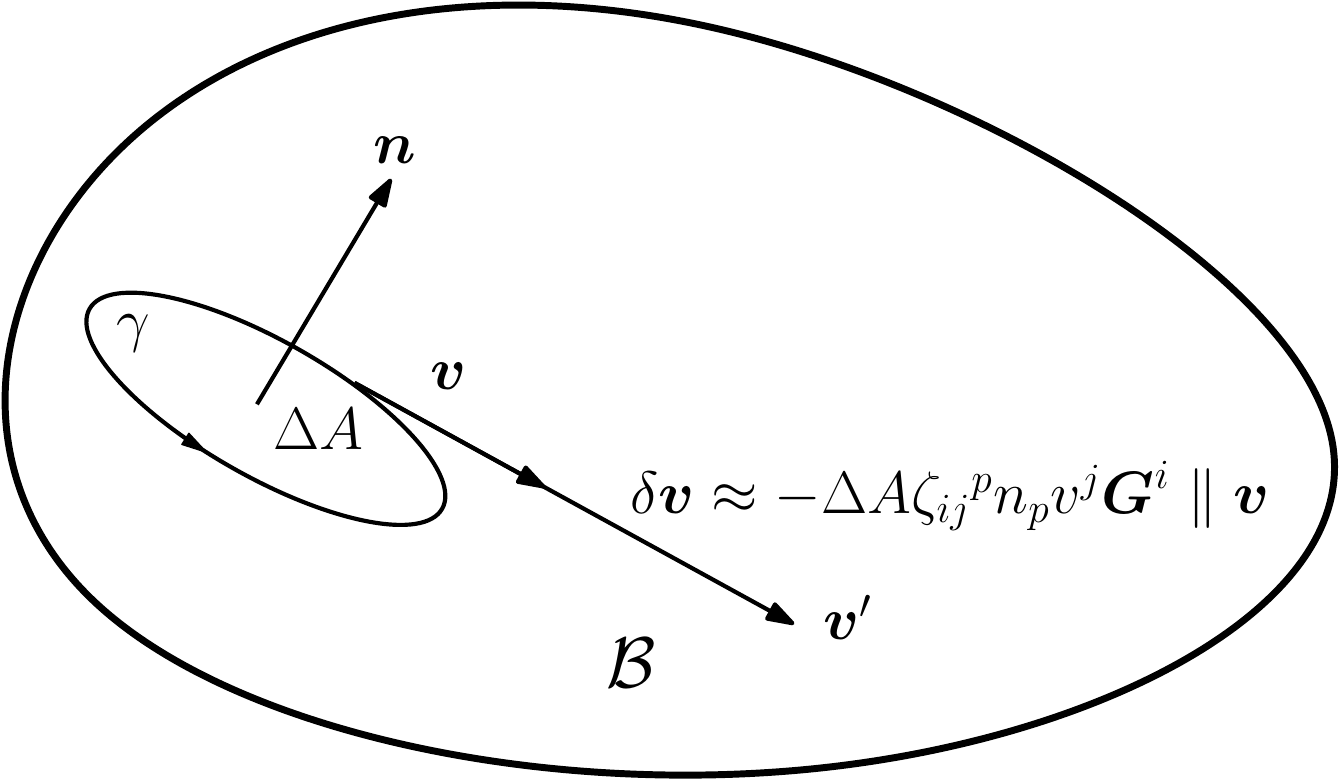}}
 \subfigure[]{\includegraphics[scale=0.55]{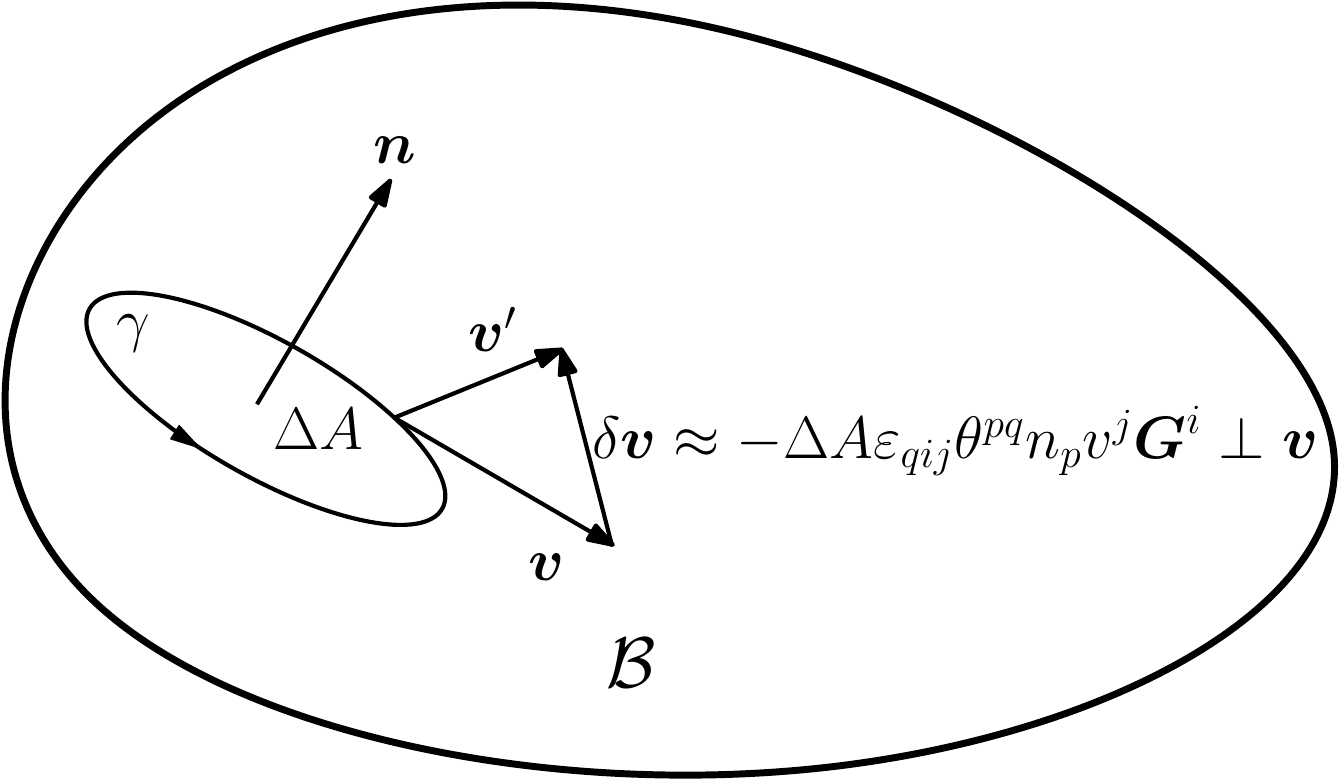}}
 \caption{(a) The symmetric part of $\mathfrak{R}(\cdot,\cdot)$, characterized by the tensor $\boldsymbol{\zeta}$, measures the stretching part of the change brought about by $\mathfrak{R}$. Here, $\boldsymbol{v}$ is a principal vector of $\boldsymbol{\zeta n}$. (b) The skew part of $\mathfrak{R}(\cdot,\cdot)$, characterized by the tensor $\boldsymbol{\theta}$, measures the purely rotational part of the change brought about by $\mathfrak{R}$. }
 \label{curvature}
\end{figure}

Let us consider an infinitesimal planar loop $\gamma$ inside $\mathcal{B}$ originating and terminating at $\boldsymbol{X}$. A tangent vector $\boldsymbol{v}$ at $\boldsymbol{X}$, when parallelly transported along $\gamma$, suffers a change $\Delta \boldsymbol{v}$ which, in the linear approximation, can be characterized by a second order tensor $\boldsymbol{\beta}:=\beta_{ij}\boldsymbol{G}^i\otimes\boldsymbol{G}^j$, i.e.,  $\Delta\boldsymbol{v}=\boldsymbol{\beta v}$, where $\beta_{ij}$ is given by \cite{schouten}
\begin{eqnarray}
 \beta_{ij}&=&-\frac{\Delta A}{2}R_{klij}\varepsilon^{rkl}\,n_r
       = -\Delta A\,\varepsilon_{qij}\, \theta^{pq}\, n_p-\Delta A\,\zeta_{ij}{}^p\,n_p.\label{change}
\end{eqnarray}
Here, $\Delta A$ is the area of the infinitesimal flat surface bounded by $\gamma$ and $\boldsymbol{n}:=n_r\boldsymbol{G}^r$ its unit normal. The first term $W_{ij}:=-\Delta A\,\varepsilon_{qij}\, \theta^{pq}\, n_p$ in the above expression is skew with an axial vector $w^m=\theta^{pm}\,n_p\,\Delta A$. It represents the amount of rotation with respect to the axis $\boldsymbol{G}_p$, for a fixed $p$, given by three Euler angles $\theta^{pq}$. Hence, $\boldsymbol{\theta}$ is the measure of a small rotation about the axis $\boldsymbol{n}$. The second term $S_{ij}:=-\Delta A\,\zeta_{ij}{}^p\,n_p$ is symmetric; it represents a stretching, with the three principal values of the tensor $\boldsymbol{\zeta n}$ as measures of the stretch along its three principal directions (Figure \ref{curvature}(a,b)).

As an example, let us assume first that $\boldsymbol{\zeta}=\boldsymbol{0}$ and the coordinates $\theta^i$ are orthonormal (i.e., $\boldsymbol{G}^i=\boldsymbol{G}_i$) locally at a point $\boldsymbol{X}$.  Let the infinitesimal loop $\gamma$ be such that $\boldsymbol{n}(\boldsymbol{X})=\boldsymbol{G}_3$. Then, $\beta_{ij}=-\Delta A\,\varepsilon_{qij}\, \theta^{3q}$. Choose $\boldsymbol{v}=\boldsymbol{G}_1$. The deviation, after parallel transport along $\gamma$, is given by $\Delta\boldsymbol{v}=-\Delta A\,\varepsilon_{qi1}\, \theta^{3q}\,\boldsymbol{G}^i=\Delta A\, \theta^{33}\boldsymbol{G}_2-\Delta A\, \theta^{32}\boldsymbol{G}_3$. Since $\Delta\boldsymbol{v}$ has no component along $\boldsymbol{v}$, it is evident that $\boldsymbol{v}$ has suffered a rotation characterized by $\theta^{32}$ and $\theta^{33}$. Next, assume that $\boldsymbol{\theta}=\boldsymbol{0}$, $\theta^i$s and $\gamma$ as above, and $v^i$ as the principal direction of $\zeta_{ij}{}^p n_p=\zeta_{ij}{}^3$ with the principal value $\lambda$, i.e., $\zeta_{ij}{}^3 v^{j}=\lambda v_i$. The deviation after parallel transport along $\gamma$ is now given by $\Delta\boldsymbol{v}=-\Delta A\,\zeta_{ij}{}^3 v^j\boldsymbol{G}^i=-\Delta  A\,\lambda\boldsymbol{v}$. Clearly, there is a stretching of the vector along its original direction.

\subsubsection*{Curvature inhomogeneities} Curvature inhomogeneities are known as disclinations. As is evident from the above discussion, there are two independent sources that might lead to disclinations: the second order tensor $\boldsymbol{\theta}$ and the third order tensor $\boldsymbol{\zeta}$.
The $\boldsymbol{\theta}$-disclinations are pure rotational anomalies in the material structure. Identification of rotational disclinations with the tensor $\boldsymbol{\theta}$ was first made by Anthony~\cite{anth70}. Rotational disclinations are the second kind of fundamental line defects which our material structure (i.e., materially uniform simple elastic solid) allows.\footnote{This is with reference to Weingarten's classical theorem \cite{weingarten1901} in linear elasticity and the subsequent construction of elementary dislocations and disclinations by Volterra \cite{volterra1907} as the fundamental line singularities in a linear elastic solid. The same construction also holds in non-linear elasticity, cf. \cite[Chapter 1]{zubov:book} and \cite{yavari-comp}. \label{fn1}} They are associated with the rotational symmetry group $\mathcal{G}$ of the material. The pure rotation that a vector suffers under parallel transport along a loop in the material space due to the presence of $\boldsymbol{\theta}$-disclination lines piercing this loop necessarily belongs to $\mathcal{G}$.

A non-zero $\boldsymbol{\zeta}$ measures the distribution of another kind of disclinations in the material structure. The disclinations characterized by the tensor $\boldsymbol{\zeta}$ are {\it not} fundamental line defects in the present material structure under consideration (see Footnote \ref{fn1}). As already seen, these are related to the stretching of vectors under parallel transport along loops, and hence, are associated with the metrical properties of the material space. We simply call them $\boldsymbol{\zeta}$-disclinations or {\it metrical disclinations}. We will shortly prove that metrical disclinations cannot exist in a metric compatible manifold (they require a certain kind of non-metricity to exist). Materials with more enriched symmetry groups, for which {\it generalized} Volterra processes exist, can indeed sustain these metrical disclinations as fundamental line defects, as has been observed in the context of general relativity \cite{kohler95}.

The absence of disclinations whatsoever is classically known as \textit{distant material parallelism}. Under distant material parallelism, crystallographic vector fields can be unambiguously defined over the whole material space. Non-zero values of either $\boldsymbol{\theta}$ or $\boldsymbol{\zeta}$ will lead to deviation from distant material parallelism. For an unambiguous definition of crystallinity at every point, distant material parallelism is a required condition (see also Section \ref{sec3}).

\subsection{Material non-metricity tensor}
The (third order) non-metricity tensor $\mathfrak{Q}:=Q_{kij}\boldsymbol{G}^i\otimes\boldsymbol{G}^j\otimes\boldsymbol{G}^k$ of the material manifold is defined as the negative of the covariant derivative of the material metric $\boldsymbol{g}$ with respect to the material connection $\mathfrak{L}$ \cite{schouten}:
\begin{equation}
  Q_{kij}:=-\nabla_k g_{ij}=-g_{ij,k}+L^p_{kj} g_{ip}+L^p_{ki} g_{jp},
  \label{def:nonmetricity}
\end{equation}
where the negative sign in the definition is conventional. It measures how the measuring scale for length and angle, i.e., the material metric, varies over the material space. It forms the third fundamental geometrical object (see Figure ~\ref{torsion&non-metricity}(b)). 
\subsubsection*{Unambiguous definition of a metric tensor field}
Let us consider an infinitesimal parametric loop $\mathcal{C}:=\{C^i(s)\boldsymbol{G}_i (s) \in \mathcal{B}\,\big|\,s\in[0,1],\,C^k(0)=C^k(1)\}$, starting and ending at the origin of the coordinate system $\theta^i$, i.e., $C^k(0)=0$, and let the metric tensor at the base point $s=0$ of this loop be given as some appropriate functions $g^0_{ij}$. We would like to investigate whether $g^0_{ij}$ remains invariant under parallel transport along the loop and, consequently, gives rise to a metric tensor field on the material space. To proceed, we transport $g^0_{ij}$ along the loop by solving the PDE
\begin{equation}
  g_{ij,k}-L^p_{kj} g_{ip}-L^p_{ki} g_{jp}=-Q_{kij}
\end{equation}
along $\mathcal{C}$, with known functions $Q_{kij}$ and $L^p_{ij}$. The above PDE follows from the definition \eqref{def:nonmetricity}. At an arbitrary position $s$ on the loop we have
\begin{equation}
 g_{ij}(s)=g^0_{ij}-\int^s_0\,Q_{kij}(\tau)\,\dot C^k(\tau)\,d\tau + \int^s_0 \, L^p_{ki}(\tau)\,g_{pj}(\tau)\,\dot C^k(\tau)\,d\tau + \int^s_0 \, L^p_{kj}(\tau)\,g_{ip}(\tau)\,\dot C^k(\tau)\,d\tau.
\end{equation}
Let us expand $Q_{kij}(\tau)$ and $g_{ij}(\tau)$ within first order in $C^k(\tau)$ as
\begin{subequations}
\begin{align}
 Q_{kij}(\tau) &\approx Q_{kij}(0) + Q_{kij,m}(0)\,C^m(\tau),\label{approx0a}\\
 g_{ij}(\tau) &\approx g^0_{ij} -Q_{kij}(0)\,C^k(\tau) +L^p_{kj}(0) g^0_{ip}\,C^k(\tau)+L^p_{ki}(0) g^0_{jp}\,C^k(\tau).
\end{align}
\label{approx0}%
\end{subequations}
After some careful calculations, it can be shown that the above relations give us the final value $g^F_{ij}$ after the parallel transport along $\mathcal{C}$ to be
\begin{equation}
 g^F_{ij}\approx g^0_{ij}+\bigg[R_{mk(ij)}-\big( Q_{kij,m} + L^p_{ki}Q_{mpj} + L^p_{kj}Q_{mpi} \big)_{[mk]} \bigg] (0)\,\oint_{\mathcal{C}}C^m\,dC^k.
\end{equation}
The expression in the square bracket is identically zero from the third Bianchi-Padova relation \eqref{diff:31}. Hence, the definition of the metric tensor is always unambiguous (path independent) in spite of the presence of non-metricity in the space, and this is the very reason why the material metric exists. But the inner product which comes via this unambiguous metric is not unambiguous, as we will see next.
\subsubsection*{Parallel transport of the inner product and its path dependence}
Let us calculate how the inner product $\boldsymbol{g}(\boldsymbol{u},\boldsymbol{v})=g_{ij}u^i v^j$ of two tangent vectors, with components $u^i$ and $v^j$, changes under parallel transport along the small parametric loop $\mathcal{C}$ as defined above. According to the definition \eqref{def:nonmetricity}, upon parallel transport to a generic point $s$ on $\mathcal{C}$, the inner product between the said vectors is given by
\begin{eqnarray}
 g_{ij}u^iv^j(s) &=& g_{ij}u^iv^j(0) + \int^s_{0} (g_{ij}u^iv^j)_{,k} (\tau) \, \dot{C}^k(\tau)\,d\tau \nonumber\\
 &=& g_{ij}u^iv^j(0) + \int^s_{0} \bigg[(g_{ij,k}u^iv^j+g_{ij} u^i_{,k} v^j +g_{ij} u^i v^j_{,k}\bigg] (\tau) \, \dot{C}^k(\tau)\,d\tau \nonumber\\
 &=& g_{ij}u^iv^j(0) + \int^s_{0} \bigg[ \big(-Q_{kij} + L^p_{ki} g_{pj}+ L^p_{kj} g_{ip}\big) u^i v^j \nonumber\\
 && + g_{ij} (\nabla_k u^i -L^i_{kp} u^p) v^j + g_{ij} u^i (\nabla_k v^j -L^i_{kp} v^p) \bigg] (\tau) \, \dot{C}^k(\tau)\,d\tau \nonumber\\
 &=& g_{ij}u^iv^j(0) - \int^s_{0} Q_{kij}(\tau) u^i(\tau) v^j (\tau)\, \dot{C}^k(\tau)\,d\tau \label{inner1}
\end{eqnarray}
since $\nabla_k u^i \,\dot C^k \equiv 0$ and $\nabla_k v^j\,\dot C^k \equiv 0$ throughout $\mathcal{C}$. This yields the well-known result (cf. \cite{steinmann}) that
{\it
inner product of arbitrary tangent vectors on a non-metric space is preserved under parallel transport if and only if non-metricity $\mathfrak{Q}$ vanishes identically.}


Let us now expand $u^i(\tau)$ and $v^j (\tau)$ around $\tau=0$, within first order in $C^k(\tau)$, as
\begin{subequations}
\begin{align}
 u^i(\tau) &\approx u^i(0) - L^i_{mp}(0) u^p(0) C^m(\tau)\,\,\,\,\textrm{and}\\
 v^j(\tau) &\approx v^j(0) - L^j_{mq}(0) v^q(0) C^m(\tau),
\end{align}
\label{approx}%
\end{subequations}
keeping in mind that the fields $u^i(\tau)$ and $v^j(\tau)$ are materially parallel. Using the approximations \eqref{approx0a} and \eqref{approx} into \eqref{inner1}, we obtain, within second order in $C^m(\tau)$,
\begin{eqnarray}
 g_{ij}u^iv^j(s) &\approx& g_{ij}u^iv^j(0) -  Q_{kij}u^iv^j\bigg|_{\tau=0}\,\int^s_{0} \dot{C}^k(\tau)\,d\tau\nonumber\\
 &&- u^iv^j\bigg|_{\tau=0}\bigg[ Q_{kij,m}(0)-L^p_{mi}(0)Q_{kjp}(0)-L^p_{mj}(0)Q_{kip}(0) \bigg]\,\int^s_{0} C^m(\tau)\dot{C}^k(\tau)d\tau. \label{inner}
\end{eqnarray}
Hence, for the loop $\mathcal{C}$,
\begin{equation}
 g_{ij}u^iv^j\bigg|_F \approx g_{ij}u^iv^j\bigg|_I - u^iv^j\bigg|_I\bigg[ Q_{kij,m}-L^p_{mi}Q_{kjp}-L^p_{mj}Q_{kip} \bigg]_I\,\oint_{\mathcal{C}} C^m(\tau)\,dC^k,
\end{equation}
where the subscripts $F$ and $I$ denote the final and initial values of the respective expressions. Since $\oint_{\mathcal{C}} C^m(\tau)\,dC^k=-\oint_{\mathcal{C}} C^k(\tau)\,dC^m$, only the skew part of the expression within the square bracket with respect to the indices $mk$ appears in the above expression, i.e.,
\begin{equation}
 g_{ij}u^iv^j\bigg|_F \approx g_{ij}u^iv^j\bigg|_I - u^iv^j\bigg|_I\bigg[ Q_{kij,m}-L^p_{mi}Q_{kjp}-L^p_{mj}Q_{kip} \bigg]_{[mk]\,\,I}\,\oint_{\mathcal{C}} C^m(\tau)\,dC^k,
\end{equation}
which can be rewritten as
\begin{equation}
 g_{ij}u^iv^j\bigg|_F \approx g_{ij}u^iv^j\bigg|_I - u^iv^j\bigg|_I\bigg[ Q_{kij,m}+L^p_{ki}Q_{mjp}+L^p_{kj}Q_{mip} \bigg]_{[mk]\,\,I}\,\oint_{\mathcal{C}} C^m(\tau)\,dC^k.
\end{equation}
Consequently, due to non-vanishing of the expression within the square bracket for a general non-metricity tensor $Q_{kij}$, parallel transport of the inner product depends on the path. For path independence of the inner product, vanishing of this expression is necessary and sufficient. Also, as we will see shortly, the third Bianchi-Padova relation implies that this expression is directly proportional to the tensor $\zeta_{ij}{}^k$ (see \eqref{diff:31}). Hence, we have
\begin{prop}\label{piq}
Inner product of arbitrary tangent vectors on a non-metric space is path independent under parallel transport if and only if
\begin{equation}
 \bigg[ Q_{kij,m}+L^p_{ki}Q_{mjp}+L^p_{kj}Q_{mip} \bigg]_{[mk]}=0
 \label{pi}
\end{equation}
identically. This condition is equivalent to the vanishing of $\boldsymbol{\zeta}$ identically over $\mathcal{B}$.
\end{prop}

In particular, a tangent vector in the material space, under parallel transport along loops, does not change its length if and only if $\boldsymbol{\zeta}$, i.e., the distribution of metrical disclinations, vanishes identically.
\subsubsection*{Non-metric inhomogeneities} The non-metricity tensor of the material space measures the non-uniformity of the material metric over the body manifold, thus, quantifying the density of a variety of metric anomalies: $(i)$ Point defects (intrinsic interstitials, vacancies, point stacking faults etc.) change the local notion of length by distorting the lattice spacings, hence, are naturally identifiable with the material non-metricity; $(ii)$ Non-uniform thermal strain or bulk material growth may inflate/deflate/shear volume elements in the material and, hence, associable with material non-metricity; $(iii)$ Magnetostrictive strain locally changes the orientation of the magnetization vector and can be associated to the non-metricity tensor in ferromagnetic materials, cf. \cite{anth71}. However, a distribution of foreign interstitials will fall in the realm of materially non-uniform bodies and is outside the scope of this work. For a treatment of such materials, see Epstein and de Le\'on \cite{epsleon00}.

\subsection{Compatibility of the geometric objects in material space}
The functions $R_{klj}{}^i$, $T_{ij}{}^k$ and $Q_{kij}$ associated with the material space cannot be arbitrary due to geometric restrictions. They satisfy the following algebraic and differential compatibility conditions (the differential conditions are known as Bianchi-Padova relations) \cite{schouten}:
\begin{subequations}
 \begin{align}
  R_{(ij)q}{}^p=0,\,\,\,& Q_{k[ij]}=0,\,\,\,T_{(ij)}{}^k=0,\label{alg}\\
  2\nabla_{[i}T_{jk]}{}^l&=R_{[ijk]}{}^l+4T_{[ij}{}^p\,T_{k]p}{}^l,\label{diff:1}\\
  \nabla_{[i}R_{jk]l}{}^p&=2 T_{[ij}{}^q\,R_{k]ql}{}^p\,\,\,\textrm{and}\label{diff:2}\\
  \nabla_{[i}Q_{j]kl}&= R_{ij(kl)}-T_{ij}{}^p\,Q_{pkl}.\label{diff:3}
 \end{align}
\label{identities}%
\end{subequations}
In the above expressions, anti-symmetrization with respect to three indices is defined as
\begin{equation}
 A_{[nml]\cdots}{}^{\cdots}:=\frac{1}{6}(A_{nml\cdots}{}^{\cdots}+A_{lnm\cdots}{}^{\cdots}+A_{mln\cdots}{}^{\cdots}-A_{lmn\cdots}{}^{\cdots}-A_{nlm\cdots}{}^{\cdots}-A_{mnl\cdots}{}^{\cdots}).
\end{equation}
The algebraic conditions \eqref{alg} follow immediately from the definitions \eqref{def:curvature}, \eqref{def:torsion} and \eqref{def:nonmetricity}. The first Bianchi-Padova relation \eqref{diff:1} is obtained by alternation of the indices $jiq$ in \eqref{def:curvature}. The second relation \eqref{diff:2} is the first order integrability condition of \eqref{def:curvature}, considered as a PDE in $L^i_{jk}$, given the functions $R_{jiq}{}^p$, whereas the last relation \eqref{diff:3} follows from the formula for the second skew covariant derivative of the material metric, i.e., $\nabla_{[i} \nabla_{j]} g_{kl}$.
Schouten \cite{schouten} has mentioned another identity where the term $R_{ijkl}-R_{klij}$ can be written in terms of the covariant derivatives of the torsion and the non-metricity tensor. In case of a Riemannian manifold, i.e., when $T_{ij}{}^p\equiv 0$ and $Q_{kij}\equiv 0$, the identities satisfied by the curvature tensor are $R_{(ij)kl}=0$, $R_{ij(kl)}=0$, $R_{[ijk]l}=0$ and $R_{ijkl}=R_{klij}$. Hence, for a Riemannian manifold, $\boldsymbol{\theta}$ is symmetric and $\boldsymbol{\zeta}$ vanishes identically.

\subsubsection*{Conservation of material inhomogeneities}
The Bianchi-Padova relations can be written in terms of various defect density tensors, namely, the disclination densities $\theta^{ij}$ and $\zeta_{ij}{}^p$, the dislocation density $\alpha^{ij}$ and the density of metric anomalies $Q_{kij}$, as (cf. \cite{povs})
\begin{subequations}
 \begin{align}
  \nabla_i \alpha^{ik} &= \varepsilon^{kmn}\theta_{nm} + \varepsilon_{ijm}\alpha^{ij}\alpha^{mk} + \zeta_{m}{}^{mk} + \frac{1}{2}\alpha^{mk} Q_{mn}{}^n,\\
  \nabla_i \theta^{ik} &= \varepsilon_{ijm} \alpha^{ij} \theta^{mk} + \frac{1}{2} \theta^{mk} Q_{mn}{}^n + \frac{1}{2} \theta^{mn} Q_{mn}{}^k - \frac{1}{2}\varepsilon^{kmn}\zeta_{mi}{}^p Q_{pn}{}^i,\\
  \nabla_i \zeta_{kr}{}^i &= \varepsilon_{ijm} \alpha^{ij} \zeta_{kr}{}^m + \frac{1}{2}\zeta_{kr}{}^m Q_{mn}{}^n + \theta^{mn} Q_{m(k}{}^p\varepsilon_{r)np} - Q_{m(k}{}^p \zeta_{r)p}{}^m\,\,\,\textrm{and}\\
  \varepsilon^{ijk}\nabla_i Q_{jmn} &= 2 \zeta_{mn}{}^k - \alpha^{kp} Q_{pmn}.
 \end{align}
\label{conservation1}%
\end{subequations}
These relations represent the conservation laws for various defect densities. Clearly, in the absence of metric anomalies, $\boldsymbol{\zeta}=\boldsymbol{0}$ and the above relations reduce down to the conservation laws for the dislocation density $\alpha^{ij}$ and the disclination density $\theta^{ij}$:
\begin{subequations}
 \begin{align}
  \nabla_i \alpha^{ik} &= \varepsilon^{kmn}\theta_{nm} + \varepsilon_{ijm}\alpha^{ij}\alpha^{mk}\,\,\,\textrm{and}\\
  \nabla_i \theta^{ik} &= \varepsilon_{ijm} \alpha^{ij} \theta^{mk}.
 \end{align}
\label{conservation2}%
\end{subequations}
Assuming $\alpha^{ij}$ and $\theta^{ij}$ to be small and of the same order, we recover the well-known conservation laws
\begin{subequations}
 \begin{align}
  \nabla_i \alpha^{ik} &= \varepsilon^{kmn}\theta_{nm} \,\,\,\textrm{and}\\
  \nabla_i \theta^{ik} &= 0,
 \end{align}
\label{conservation3}%
\end{subequations}
which state the classical result that in absence of metric anomalies, dislocation lines must end on disclinations (within the body) and disclination lines cannot end inside the body \cite{anth70}.

\section{Representation of metric anomalies}\label{sec3}
For anisotropic elastic solids (e.g., crystalline materials), the (rotational) symmetry group $\mathcal{G}$ is discrete. In the continuum limit of such a material from its discrete state, the translational symmetry parameters, which have the order of magnitude of one lattice parameter, get infinitesimally small. But the rotational symmetries (the elements of $\mathcal{G}$) remain finite in the continuum limit. The rotation that a tangent vector suffers under parallel transport along a loop in the material space that encircles the $\boldsymbol{\theta}$-disclination lines necessarily belongs to the symmetry group $\mathcal{G}$. If the loop encircles infinite number of these disclination lines, as is the case when a continuous distribution of disclinations is present, the resulting rotational deviation can go unbounded, leading to an unbounded elastic energy. Due to such unrealistic energy penalty, a continuous distribution of $\boldsymbol{\theta}$-disclinations has never been observed in crystalline materials having discrete rotational symmetry groups. It is therefore reasonable to assume $\boldsymbol{\theta}=\boldsymbol{0}$ in anisotropic simple elastic materials \cite{anth02}. Moreover, as the metrical disclinations lead to ambiguity in the definition of inner product field and are not fundamental line defects within the scope of the present class of materials being considered, we will also assume $\boldsymbol{\zeta}$ to be identically zero.

Under this assumption of absolute distant material parallelism, we will now discuss three possible models of the non-metricity tensor in the following.

\subsection{Irrotational metric anomalies} \label{irrotma}
The third Bianchi-Padova relation \eqref{diff:3} can be rewritten as
\begin{equation}
 \big(Q_{jkl,i}+L^p_{jk}Q_{ipl}+L^p_{jl}Q_{ipk}\big)_{[ji]}= R_{ij(kl)}
 \label{diff:31}
\end{equation}
or equivalently
\begin{equation}
 \frac{1}{2}\varepsilon^{qij}\big(Q_{jkl,i}+L^p_{jk}Q_{ipl}+L^p_{jl}Q_{ipk}\big)=\zeta_{kl}{}^q.
 \label{diff:32}
\end{equation}
Clearly, $Q_{kmj}=0$ is the trivial solution of the last equation for $\zeta_{km}{}^p=0$, but there exist non-trivial solutions. For $\boldsymbol{\theta} = \boldsymbol{0}$, we seek the most general form of the non-metricity tensor that uniquely corresponds to $\boldsymbol{\zeta}=\boldsymbol{0}$. In the absence of $\boldsymbol{\theta}$-disclinations, such a form of non-metricity is necessary to maintain absolute distant material parallelism throughout the body. We have

\begin{prop}\label{uni}
 If $\mathcal{B}$ is simply connected and $\boldsymbol{\theta}=\boldsymbol{0}$, then the necessary condition for $\boldsymbol{\zeta}=\boldsymbol{0}$ is that there exists a tensor field $\boldsymbol{q}:=q_{ij}\boldsymbol{G}^i\otimes\boldsymbol{G}^j:\mathcal{B}\to Sym$ such that $Q_{kij}=-2\nabla_k q_{ij}$. Moreover, if $\boldsymbol{q}$ is such that $\bar{\boldsymbol{g}}:=\boldsymbol{g}-2\boldsymbol{q}$ is positive definite, then the above condition is also sufficient.
\end{prop}
The factor $-2$ appearing in the form of $Q_{kij}$ is conventional (cf. \cite[Equation (33)]{anth71}). Note that the sufficient condition involves the material metric $\boldsymbol{g}$. Towards proving this proposition, we will need the following theorem. For our purpose, sufficient regularity of the respective fields can be assumed. The notations used for various functional spaces are standard.
\begin{thm}
 (Fundamental existence theorem for linear differential systems \cite{mardare03a}) Let $\Omega$ be a simply connected open subset of $\mathbb{R}^p$ whose geodesic diameter is finite, and let $q\ge 1$ be an integer. Let there be matrix fields $A_\alpha\in L^\infty(\Omega,M^{q})$, $B_\alpha\in L^\infty(\Omega, M^{p})$ and $C_\alpha\in L^\infty(\Omega, M^{p\times q})$ such that
 \begin{subequations}
  \begin{align}
   A_{\alpha,\beta}+A_\alpha A_\beta &= A_{\beta,\alpha} + A_\beta A_\alpha,\label{cond1}\\
   B_{\alpha,\beta}+B_\alpha B_\beta &= B_{\beta,\alpha} + B_\beta A_\alpha \,\,\,\textrm{and}\label{cond2}\\
   C_{\alpha,\beta}+C_\beta A_\alpha + B_\alpha C_\beta &= C_{\beta,\alpha} + C_\alpha A_\beta + B_\beta C_\alpha\label{cond3}
  \end{align}
  \label{cond}%
 \end{subequations}
are satisfied in $\mathcal{D}'(\Omega,M^q)$, $\mathcal{D}'(\Omega,M^{p})$ and $\mathcal{D}'(\Omega,M^{p\times q})$, respectively. In this theorem, the Greek indices $\alpha$, $\beta$ varies over $1,2,\ldots,p$. Then there exists a matrix field $Y\in W^{1,\infty}(\Omega,M^{p\times q})$ that satisfies
 \begin{equation}
  Y_{,\alpha}=Y A_\alpha + B_\alpha Y + C_\alpha \,\,\,\,\textrm{in}\,\,\,\mathcal{D}'(\Omega,M^{p\times q}).
 \end{equation}
\end{thm}

\noindent {\bf Proof of Proposition \ref{uni}:}
 In the above theorem, choose $p=q=3$, $\Omega=\mathcal{B}$, $A_k = B_k = [L^i_{kj}]$ and $C_k = [Q_{kij}]$. Then the conditions \eqref{cond1} and \eqref{cond2} amount to $R_{ijq}{}^p=0$, i.e.~$\boldsymbol{\theta}=\boldsymbol{0}$ and $\boldsymbol{\zeta}=\boldsymbol{0}$ and the condition \eqref{cond3} yields $\big(Q_{jkl,i}+L^p_{jk}Q_{ipl}+L^p_{jl}Q_{ipk}\big)_{[ij]}=0$ which, due to \eqref{diff:32}, is equivalent to $\boldsymbol{\zeta}=\boldsymbol{0}$.  
Hence, as a consequence of the above theorem, there exists a sufficiently regular matrix field $[q_{ij}]$ such that
\begin{equation}
 Q_{kij}=-2q_{ij,k}+2L^p_{ki}q_{pj}+2L^p_{kj}q_{ip}=-2\nabla_kq_{ij}.
\end{equation}
The symmetry $Q_{kij}=Q_{k(ij)}$ implies the symmetry of the matrix field $[q_{ij}]$.

To prove the sufficiency,  we insert $Q_{kij}=-2\nabla_k q_{ij}$ in \eqref{diff:31} to obtain
\begin{equation}
 -R_{jikm} q^m{}_{l} -R_{jilm} q^m{}_{k}=R_{ij(kl)}. \label{eq37}
\end{equation}
Here, $q^m{}_{k}:=g^{im}q_{ik}$. For the time being, let us assume that $\boldsymbol{\zeta}\ne\boldsymbol{0}$. For $\boldsymbol{\theta}=\boldsymbol{0}$, $R_{jikm}=R_{ji(km)}=\varepsilon_{pji}\zeta_{km}{}^{p}$, \eqref{eq37} reduces to
\begin{equation}
 \varepsilon_{pji} (\zeta_{km}{}^p q^m{}_{l} + \zeta_{lm}{}^p q^m{}_{k})= - \varepsilon_{ijp}\zeta_{kl}{}^p,
\end{equation}
which implies
\begin{equation}
 (q_{ml}+\frac{1}{2}g_{ml}) \zeta^m{}_{k}{}^p  +  (q_{mk}+\frac{1}{2}g_{mk}) \zeta^m{}_{l}{}^p = 0.
\end{equation}
We now find out the conditions on the matrix $[q_{ij}]$ such that the $3\times 3$ matrix $[\zeta^i{}_{j}{}^p]$, for each $p$, is identically zero. Recalling that $[g_{ij}-2q_{ij}]$ is symmetric, there exists a basis in which it is a diagonal matrix. In that basis, $[g_{ij}-2q_{ij}]$ can be written as $\mbox{diag}(a_1,a_2,a_3)$, where $a_i$s are the three real eigenvalues of $[g_{ij}-2q_{ij}]$. Moreover, let us denote by $[b^i_{j}]:=[\zeta^i{}_{j}{}^p]$ for some fixed $p$. The last expression boils down to
\begin{equation}
 \mbox{diag}(2a_1,a_1+a_2,a_1+a_3,a_1+a_2,2a_2,a_2+a_3,a_1+a_3,a_3+a_2,2a_3)\left[\begin{array}{c}
                                                                                   b^1_{1}\\
                                                                                   b^1_{2}\\
                                                                                   b^1_{3}\\
                                                                                   b^2_{1}\\
                                                                                   b^2_{2}\\
                                                                                   b^2_{3}\\
                                                                                   b^3_{1}\\
                                                                                   b^3_{2}\\
                                                                                   b^3_{3}
                                                                                  \end{array}
\right]=\left[\begin{array}{c}
                                                                                   0\\
                                                                                   0\\
                                                                                   0\\
                                                                                   0\\
                                                                                   0\\
                                                                                   0\\
                                                                                   0\\
                                                                                   0\\
                                                                                   0
                                                                                  \end{array}
\right].
\label{condition_on_q}
 \end{equation}
The condition for $[b^i_{j}]$, hence $\boldsymbol{\zeta}$, to vanish identically is that the determinant of the $9\times 9$ diagonal matrix in the above expression is non-zero, i.e.,
\begin{equation}
 a_1a_2a_3(a_1+a_2)(a_2+a_3)(a_3+a_1)\ne 0.
 \label{eq}
\end{equation}
Hence, if $[q_{ij}]$ is such that the eigenvalues $a_i$s of $[g_{ij}-2q_{ij}]$ satisfies \eqref{eq} everywhere, then $\boldsymbol{\zeta}$ vanishes identically. For a positive definite $[\bar{g}_{ij}]:=[g_{ij}-2q_{ij}]$, \eqref{eq} is always satisfied.\hfill$\square$


\begin{figure}[t!] 
\centering
\subfigure[]{\includegraphics[scale=0.4]{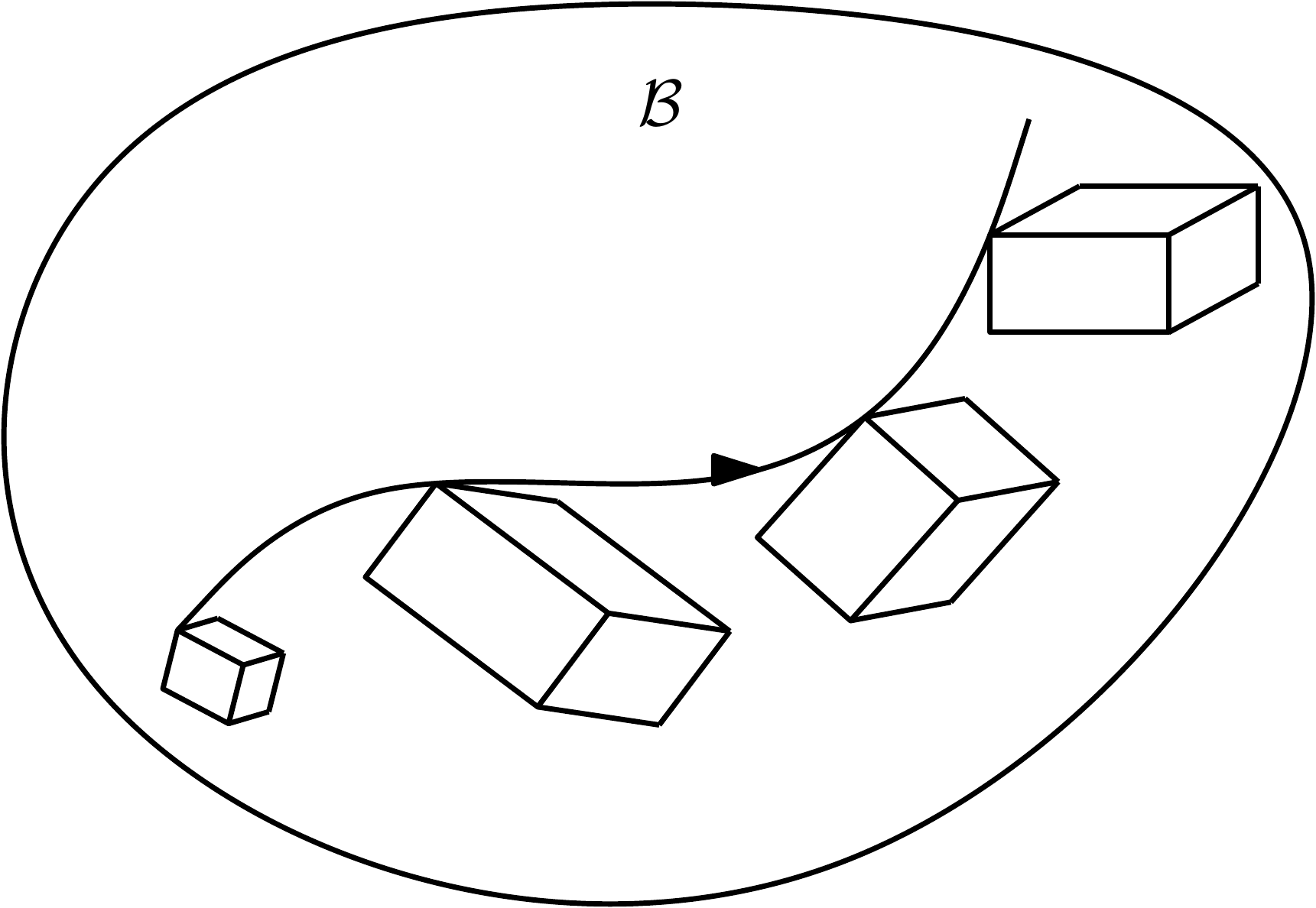}}
\hspace{15mm}
\subfigure[]{\includegraphics[scale=0.4]{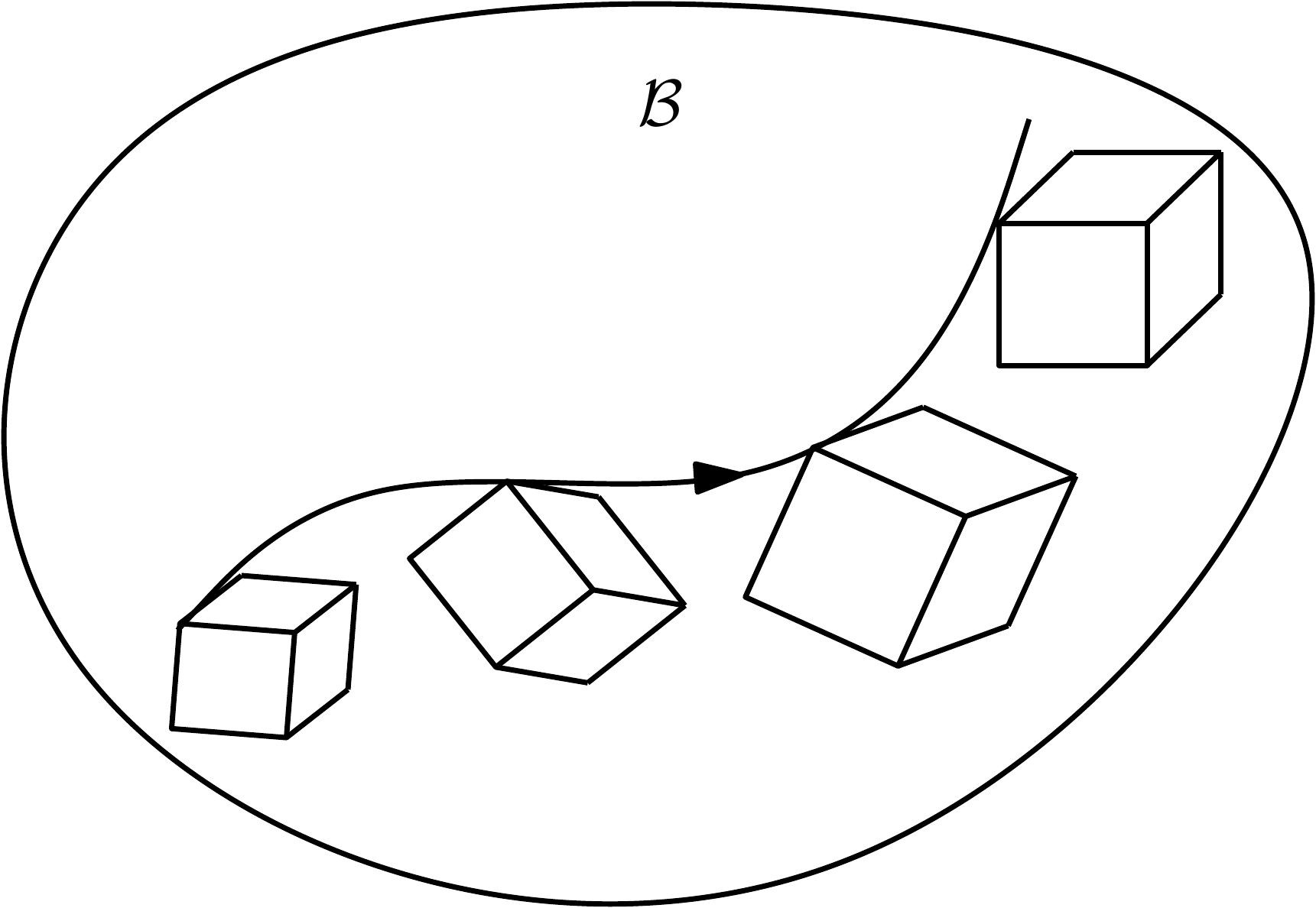}}
\caption{(a) General irrotational metric anomalies change the orientation, shape and size of a cube along a curve, whereas (b) isotropic metric anomalies, which are also irrotational, change the orientation and size, but not the shape, of a cube along a curve.}
\label{irrot_and_iso}
\end{figure}
Metric anomalies characterized by $Q_{kij}=-2\nabla_kq_{ij}$, where $\boldsymbol{q}$ satisfies the necessary and sufficient conditions of Proposition \ref{uni}, are called {\it irrotational}, because they uniquely correspond to the vanishing of material curvature. Only irrotational metric anomalies are allowed under our present assumption of absolute distant parallelism in a materially uniform simple elastic solid. The tensor field $\boldsymbol{q}$ is a complete measure of irrotaional metric anomalies. It induces a field of orthonormal triple of eigenvectors $\{\boldsymbol{a},\boldsymbol{b},\boldsymbol{c}\}$, corresponding to a field of eigenvalues $a$, $b$ and $c$, respectively: $\boldsymbol{qa}=a\boldsymbol{a}$, $\boldsymbol{qb}=b\boldsymbol{b}$ and $\boldsymbol{qc}=c\boldsymbol{c}$. The tensor field $\boldsymbol{q}$, as well as its field of eigenpairs $\{a,\boldsymbol{a}\}$ etc.~can be restricted to a curve. Then, we have a field of rectangular parallelepiped formed by the triple of orthogonal vectors $\{a\boldsymbol{a},b\boldsymbol{b},c\boldsymbol{c}\}$ over this curve (see Figure \ref{irrot_and_iso}(a)). This field of rectangular parallelepiped is uniquely defined over the whole material space because of distant material parallelism. The parallelepiped returns to its initial shape, size and orientation after circumnavigation along a loop. The formula \eqref{inner} for inner product of arbitrary vectors under parallel transport, in presence of purely irrotational metric anomalies, reduces down to (upto leading order)
\begin{eqnarray}
 g_{ij}u^iv^j(s) &\approx& g_{ij}u^iv^j(0) +2  (\nabla_kq_{ij}\,u^iv^j)\bigg|_{\tau=0}\,\int^s_{0} \dot{C}^k(\tau)\,d\tau.
 \label{irrot}
\end{eqnarray}

\subsubsection*{Quasi-plastic strain}
For irrotational metric anomalies, the positive definite symmetric tensor field $\bar{\boldsymbol{g}}$ can be used to define an {\it auxiliary material space} $(\mathcal{B},\mathfrak{L},\bar{\boldsymbol{g}})$, equipped with the original material connection $\mathfrak{L}$ and the metric $\bar{\boldsymbol{g}}$. The non-metricity  of this auxiliary material space vanishes identically by definition: $\nabla_k\bar{g}_{ij}=\nabla_k(g_{ij}-2q_{ij})=-Q_{kij}+Q_{kij}=0$. The curvature is identically zero for both $(\mathcal{B},\mathfrak{L},{\boldsymbol{g}})$ and $(\mathcal{B},\mathfrak{L},\bar{\boldsymbol{g}})$. Auxiliary material space is a geometric space derived from the material space, with identical torsion and curvature fields, whose non-metricity is identically zero.

Anthony \cite{anth70, anthony70a} (cf. \cite{falk81}) used  $\boldsymbol{q}$ (calling it {\it quasi-plastic strain}) to study metric anomalies when absolute distance parallelism is maintained throughout the material space. The fundamental geometric reasoning behind the existence of $\boldsymbol{q}$, as we discussed above, was absent in their work. The terminology ``strain'' is clear from the relation $\boldsymbol{q}=\frac{1}{2}(\boldsymbol{g}-\bar{\boldsymbol{g}})$, i.e., $\boldsymbol{q}$ is the difference between the respective metric tensors of the material space and the auxiliary material space.
\begin{rem} 
({\it Isotropic metric anomalies.}) {\rm The second order tensor $\boldsymbol{q}$ characterizing irrotational metric anomalies has the unique decomposition
\begin{equation}
 q_{ij}=\lambda g_{ij}+\mathfrak{q}_{ij},
\end{equation}
where $\lambda:=\frac{1}{3}q^k{}_{k}$ is the trace of $\boldsymbol{q}$ ($\lambda g_{ij}$ is called the spherical/isotropic part), and $\mathfrak{q}_{ij}$ is the deviatoric part of $q_{ij}$, i.e., $\mathfrak{q}^k{}_{k}=0$. Let us consider the case when $\boldsymbol{q}$ is purely isotropic, i.e., $q_{ij}=\lambda g_{ij}$. Then, it is straightforward to obtain $Q_{kij}=-\mu_{,k}g_{ij}$, where $\mu:=\ln(1+2\lambda)$. In this case, the formula \eqref{irrot} for the inner product of arbitrary vectors under parallel transport along a curve $\mathcal{C}$ reduces to
\begin{eqnarray}
 g_{ij}u^iv^j(s) &\approx& g_{ij}u^iv^j(0)\bigg[1 +  \mu_{,k} \big|_{\tau=0}\,\int^s_{0} \dot{C}^k(\tau)\,d\tau\bigg].
 \label{isotropic}
\end{eqnarray}
Hence, orthogonal vectors always remain orthogonal under parallel transport. Since all the eigenvalues of an isotropic tensor are equal, the cube formed by the eigenpairs $\{\lambda\boldsymbol{a},\lambda\boldsymbol{b},\lambda\boldsymbol{c}\}$, where $\{\boldsymbol{a},\boldsymbol{b},\boldsymbol{c}\}$ are any triple of orthonormal vectors forming the eigenspace of $\boldsymbol{q}$ (any orthonormal triple of vectors forms the eigenspace of an isotropic tensor), inflates/deflates as one moves along the curve $\mathcal{C}$ (see Figure \ref{irrot_and_iso}(b)). The auxiliary material space for isotropic metric anomalies is conformal to the original material space, because $\bar{\boldsymbol{g}}=(1+2\lambda)\boldsymbol{g}$.
} \label{isoma}
\end{rem}

\begin{rem}
({\it Regularity of the induced field of parallelepipeds.}) {\rm The one parameter fields of eigenvalues of the one parameter tensor field $\boldsymbol{q}(s)$ along a parametric curve $\mathcal{C}$ always posses the same regularity as that of $\boldsymbol{q}(s)$  \cite{laxbook, rellichbook}. If all the eigenvalues are simple throughout $\mathcal{C}$, then the corresponding field of eigenvectors has the same regularity as that of $\boldsymbol{q}$. This is also true if the multiplicity of all the eigenvalues remains constant throughout the curve. In case of isotropic metric anomalies, the multiplicity is 3 throughout $\mathcal{C}$; hence, the field of cubes has the same regularity as the function $\mu_{,k}(s)$ along $\mathcal{C}$.
 }
\end{rem}


\subsection{Semi-metric geometry}
For a second representation of metric anomalies we look into semi-metric geometry. In semi-metric geometry, $Q_{kmj}$ is given in terms of a vector $Q_k$ as $Q_{kmj}=Q_k g_{mj}$ (semi-metric geometry with zero torsion is called Weyl-geometry) \cite{schouten}. This form of $Q_{kij}$, when plugged into the third Bianchi-Padova relation \eqref{diff:31}, reduces it to
 \begin{equation}
  Q_{[j,i]}g_{km}=R_{ij(km)}.
  \label{semim}
 \end{equation}
If $\boldsymbol{\theta}=\boldsymbol{0}$ and the domain $\mathcal{B}$ is simply connected, \eqref{semim} implies, from Poincar\'e's lemma, that $R_{ij(km)}=0$ if and only if there exists a function $\phi:\mathcal{B}\to \mathbb{R}$ such that $Q_i=\phi_{,i}$  \cite{gunther}. Define $\psi:=\exp \phi -1$. Then, $\phi_{,i}=\frac{\psi_{,i}}{\psi+1}$; hence, $\nabla_k \bar{g}_{ij}=0$ where $\bar{g}_{ij}:=(\psi+1) g_{ij}$, i.e., we recover the isotropic representation discussed in Remark \ref{isoma}. Moreover, using the Helmholtz representation (decomposition of a vector field into a curl free and a divergence free part) of the vector field $Q_j$ as  $Q_j=\phi_{,j}+g_{ij}\varepsilon^{imn}\nabla_m \mathfrak{q}_{n}$, with $\nabla_n \mathfrak{q}_{n}=0$ (cf. \cite{hehl}), we have the following
\begin{prop}\label{semimetric}
 In semi-metric geometry, with $\mathcal{B}$ simply connected and $\boldsymbol{\theta}=\boldsymbol{0}$, the curl free part of $Q_{k}$, expressed as $\phi_{,k}$ for some scalar function $\phi$, and the divergence free part of $Q_{k}$, characterized by the vector field $\mathfrak{q}_i$ as above, uniquely correspond to $\boldsymbol{\zeta}=\boldsymbol{0}$ and $\boldsymbol{\zeta}\ne\boldsymbol{0}$, respectively.
\end{prop}

The concept of non-metricity, as well as its semi-metric form (in fact, with vanishing torsion), was introduced by Hermann Weyl \cite[pp.~121-125]{weyl} in an attempt to unify gravity with electromagnetism. The semi-metric form of non-metricity preserves the ratio of the magnitude of two vectors during parallel transport along a curve. Indeed (cf. \cite{steinmann}), let $u=g_{ij}u^i u^j$ and $v=g_{ij}v^i v^j$ be the squared lengths of two tangent vectors at any point on a curve $\mathcal{C}$. Then, upon parallel transport along an infinitesimal sector $dC^k$, the changes in $u$ and $v$, by \eqref{inner1}, are given by $\Delta u=-u\,Q_k \,dC^k$ and $\Delta v=-v\,Q_k \,dC^k$, respectively. Hence,
\begin{equation}
 \Delta\bigg(\frac{u}{v}\bigg)=\frac{\Delta u}{v}-\frac{u}{v^2}\Delta v=-\bigg(\frac{u}{v}-\frac{u}{v}\bigg)Q_k\,dC^k=0.
\end{equation}

Semi-metric geometry, with $Q_i=\phi_{,i}$, has been used by Miri and Rivier \cite{miri} and more recently by Yavari and Goriely \cite{yavari12, yavari14} to model isotropic metric anomalies in the context of a distribution of spherically symmetric point defects. It is clear from Proposition \ref{semimetric} that, with $\boldsymbol{\theta}=\boldsymbol{\zeta}=\boldsymbol{0}$, the semi-metric model can represent only isotropic metric anomalies. In this context, the scope of quasi-plastic strain model discussed previously is larger and physically amenable in representing anisotropic metric anomalies.

\subsection{Quasi-plastic deformation}\label{secqpd}
\begin{figure}[t!] 
\centering 
\includegraphics[scale=0.6]{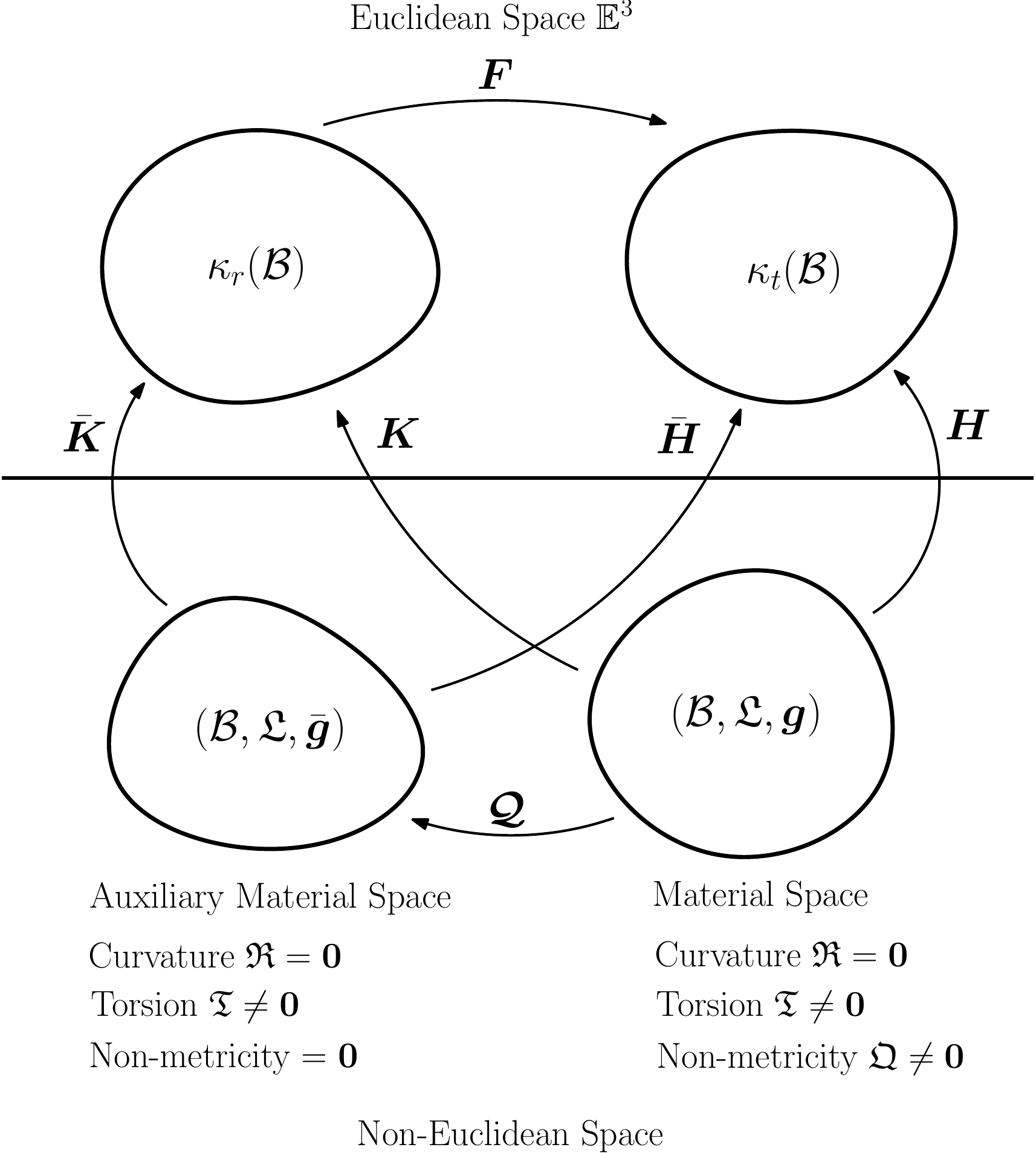}
\caption{Mappings between the tangent spaces of various configurations and spaces associated with the material manifold $\mathcal{B}$, see Section \ref{secqpd} for details.}
\label{multi}
\end{figure}

The auxiliary material space $(\mathcal{B},\mathfrak{L},\bar{\boldsymbol{g}})$, defined in Section \ref{irrotma}, inherits the affine connection $\mathfrak{L}$ (with zero curvature) from the material space $(\mathcal{B},\mathfrak{L},\boldsymbol{g})$ but has a different metric field $\bar{\boldsymbol{g}}$ such that its non-metricity vanishes identically, i.e., $\nabla_k\bar{g}_{ij}=0$. With both curvature and non-metricity identically zero, the auxiliary material space can still have non-trivial torsion. Therefore $(\mathcal{B},\mathfrak{L},\bar{\boldsymbol{g}})$ can support only dislocations as possible sources of inhomogeneity. In this scenario, according to a classical result in differential geometry (cf. \cite[Theorem 2.1]{arcgupta1}), there exists a sufficiently smooth invertible tensor field $\bar{\boldsymbol{H}}:=\bar H_{ij}\boldsymbol{G}^i\otimes\boldsymbol{G}^j$ over $\mathcal{B}$ such that
\begin{subequations}
\begin{align}
 L_{ij}^p &= (\bar H^{-1})^{pl}\bar{H}_{li,j}\,\,\,\textrm{and}\label{dd}\\
 \bar{\boldsymbol{g}} &= \bar{\boldsymbol{H}}^{T}\bar{\boldsymbol{H}}.\label{dd1}%
 \end{align}
\end{subequations}
The auxiliary metric $\bar{\boldsymbol{g}}$ is positive definite by construction. We assume $\mbox{det}~\bar{\boldsymbol{H}} > 0$. The tensor $\bar{\boldsymbol{H}}$ maps the tangent spaces of the auxiliary material space to the tangent spaces of the current configuration $\kappa_t(\mathcal{B})\subset\mathbb{E}^3$ (see Figure \ref{multi}). Here, $\mathbb{E}^3$ denotes the 3-dimensional Euclidean point space.

We assume that there exists a sufficiently smooth tensor field in $InvLin^+$, $\boldsymbol{\mathcal{Q}}:=\mathcal{Q}^i{}_{j}\boldsymbol{G}_i\otimes\boldsymbol{G}^j$, over $\mathcal{B}$ which maps the tangent spaces of the material space to the tangent spaces of the auxiliary material space, such that
\begin{equation}
  \boldsymbol{g}=\boldsymbol{\mathcal{Q}}^T\bar{\boldsymbol{g}}\boldsymbol{\mathcal{Q}}.
  \label{auxiliary_metric}
\end{equation}
The tensor $\boldsymbol{\mathcal{Q}}$ is the third representation of metric anomalies discussed in this paper. We call it {\it quasi-plastic deformation} for reasons that will be discussed below.
The preceding assumption is tantamount to the existence of a well-defined material uniformity field and a crystallographic basis field over the material space. Indeed, substituting \eqref{dd1} in \eqref{auxiliary_metric} allows us to write elastic metric as
\begin{equation}
  \boldsymbol{g}= \boldsymbol{H}^{T}\boldsymbol{H}, \label{metricnew}
\end{equation}
where $\boldsymbol{H}= \bar{\boldsymbol{H}} \boldsymbol{\mathcal{Q}}$ is a sufficiently smooth tensor field in $InvLin^+$ which maps the tangent spaces of the material space to the tangent spaces of the current configuration of the body (see Figure \ref{multi}). The tensor $\boldsymbol{H}$ is called the elastic deformation tensor. Consider a fixed reference configuration $\kappa_r(\mathcal{B}) \subset \mathbb{E}^3$ and let $\boldsymbol{F}  \in InvLin^+$ be the deformation gradient tensor which maps tangent spaces in $\kappa_r$ to those in $\kappa_t$ (see Figure \ref{multi}). The field $\boldsymbol{h}\in Sym^+$ introduced in Section \ref{uniform} is of the form $\boldsymbol{F}^T \boldsymbol{F}$. Recalling the discussion in Section \ref{uniform}, and using \eqref{metricnew}, we can construct a well-defined smooth material uniformity field $\boldsymbol{K} \in InvLin^+$ such that (see Figure \ref{multi})
\begin{equation}
\boldsymbol{H}=\boldsymbol{FK}. \label{plastmulti}
\end{equation}
The tensor $\boldsymbol{K}$ is conventionally called the plastic deformation tensor. The tensors $\boldsymbol{H}$ and $\boldsymbol{K}^{-1}$ are usually denoted as $\boldsymbol{F}^e$ and $\boldsymbol{F}^p$, respectively, in the plasticity literature. With the above mappings in place, we can define an auxiliary plastic deformation tensor $\bar{\boldsymbol{K}} \in InvLin^+$ such that $\bar{\boldsymbol{H}}=\boldsymbol{F} \bar{\boldsymbol{K}}$. The multiplicative decomposition 
\begin{equation}
\boldsymbol{K}= \bar{\boldsymbol{K}} \boldsymbol{\mathcal{Q}}
 \label{qpd}
\end{equation}
of the plastic deformation tensor follows immediately (see Figure \ref{multi}). With the existence of $\boldsymbol{\mathcal{Q}}$ we can also construct an unambiguous crystallographic vector field $\boldsymbol{g}_i=\boldsymbol{H}\boldsymbol{G}_i$ such that $g_{ij} = \boldsymbol{g}_i \boldsymbol{\cdot} \boldsymbol{g}_j$.

The motivation for introducing $\boldsymbol{\mathcal{Q}}$ is clear from the multiplicative decompositions \eqref{plastmulti} and \eqref{qpd}. 
Consider the case when the material space $(\mathcal{B},\mathfrak{L},\boldsymbol{g})$ has only metric anomalies and is therefore free of dislocations. Then, the auxiliary material space $(\mathcal{B},\mathfrak{L},\bar{\boldsymbol{g}})$ is free of any inhomogeneity, and will be a connected subset of $\mathbb{E}^3$. We can identify it with the reference configuration $\kappa_r$, i.e., $\bar{\boldsymbol{K}}= {\boldsymbol{I}}$ and $\bar{\boldsymbol{H}} = \boldsymbol{F}$ identically over the whole domain. The tensor $\boldsymbol{\mathcal{Q}}$ then determines the plastic distortion $\boldsymbol{K}$, hence the terminology quasi-plastic deformation. On the other hand, when the material space is dislocated and also has metric anomalies, the components of the torsion tensor can be calculated from \eqref{dd} as
\begin{equation}
 T_{ij}{}^p=({\bar{H}}^{-1})^{pl}{\bar{H}}_{l[i,j]}.
\end{equation}
It is clear that the information about dislocation density in the material space is contained only in the incompatibility of $\bar{\boldsymbol{H}}$ (or equivalently of $\bar{\boldsymbol{K}}$) . The tensor $\boldsymbol{\mathcal{Q}}$ can then be understood to contain information about the metric anomalies, as described in the following paragraph. The proposed framework can be seen as a generalization of a version of the fundamental theorem of Riemannian geometry in the context of continuum theory of material defects, as stated and proved in Roychowdhury and Gupta \cite[Theorem 2.1]{arcgupta1}, by including metric anomalies into consideration. In the absence of non-metricity we recover Theorem 2.1 in \cite{arcgupta1}. 

Combining equations $\nabla_k\bar{g}_{ij}=0$ and $Q_{kij}=-\nabla_k g_{ij}$ with \eqref{auxiliary_metric}, it is straightforward to relate the non-metricity tensor of the material space to the quasi-plastic deformation tensor as
\begin{equation}
 Q_{kij}=2\bigg[(\nabla_k\mathcal{Q}_{ip})\,(\mathcal{Q}^{-1})^p{}_j\bigg]_{(ij)},
  \label{nonmetricity_quasiplastic}
\end{equation}
where $\mathcal{Q}_{ij}:=g_{ip}\mathcal{Q}^p{}_j$.
However, the quasi-plastic deformation $\boldsymbol{\mathcal{Q}}$ cannot be an arbitrary tensor. According to the third Bianchi-Padova relation \eqref{diff:3}, it has to necessarily satisfy the following second order non-linear PDE in order to conform to the vanishing of $\boldsymbol{\theta}$ and $\boldsymbol{\zeta}$:
\begin{equation}
\bigg[ (\nabla_{[m}\nabla_{k]}\mathcal{Q}_{ip}^{-1})\,\,\mathcal{Q}^p{}_j
 +(\nabla_{[m}\mathcal{Q}^p{}_{|i|})\,\,\nabla_{k]}(\mathcal{Q}_{jp})^{-1}
 - T_{mk}{}^p\,(\nabla_p\mathcal{Q}_{iq})\,\,(\mathcal{Q}^{-1})^q{}_{j} \bigg]_{(ij)}=0.
 \label{cond:qpd}
\end{equation}
This equation has been obtained by substituting \eqref{nonmetricity_quasiplastic} into \eqref{diff:3} and imposing $R_{ijp}{}^q=0$. The indices enclosed within the vertical bars $|\cdot|$ are exempt from any symmetrization or anti-symmetrization operation. We should emphasize that a description of metric anomalies in terms of the quasi-plastic deformation tensor $\boldsymbol{\mathcal{Q}}$ is an alternative model for irrotational metric anomalies in crystalline solids as proposed in Section \ref{irrotma}. While the present representation allows us to obtain elegant multiplicative decompositions of total and plastic deformations, it comes at the cost of satisfying conditions \eqref{cond:qpd}. The quasi-plastic strain model in Section \ref{irrotma} is free from such constraints but provides no basis for multiplicative decompositions. Both of these representations can be used to model anisotropic metric anomalies. By comparing \eqref{auxiliary_metric} with an equation for $\bar{\boldsymbol{g}}$ in Proposition \ref{uni}, we can obtain a relation between $\boldsymbol{\mathcal{Q}}$ and $\boldsymbol{q}$:
\begin{equation}
\boldsymbol{\mathcal{Q}}^{-T}\boldsymbol{g}\boldsymbol{\mathcal{Q}}^{-1} = \boldsymbol{g} - 2 \boldsymbol{q}.
  \label{sqbq}
\end{equation}
For a given $\boldsymbol{q}$ these provide only six (nonlinear) equations to be solved for $\boldsymbol{\mathcal{Q}}$. This is illustrated clearly in the linearized setting of Remark  \ref{linearqp} below. In the absence of metric anomalies it is reasonable to take $\boldsymbol{q} = \boldsymbol{0}$ (so that $\bar{\boldsymbol{g}}$ is identical to ${\boldsymbol{g}}$) and $\boldsymbol{\mathcal{Q}} = \boldsymbol{I}$ (so that the auxiliary material space is identical to the material space).

A multiplicative decomposition framework, such as that provided by Equations \eqref{plastmulti} and \eqref{qpd}, is useful for analytical and numerical studies of displacement boundary-value-problems of inhomogeneous solids. Our purpose is to provide a rigorous setting in which such a decomposition can be justified in the presence of metric anomalies. Moreover, representation of metric anomalies by $\boldsymbol{\mathcal{Q}}$ allows us to take into account more general distortional defect densities such as those arising in a distribution of point stacking faults \cite{kroner90, kroner94}.

We summarize the above discussion in the following
\begin{prop} \label{qpdprop}
 If the non-metricity $\mathfrak{Q}$ is given in terms of the quasi-plastic deformation $\boldsymbol{\mathcal{Q}}$ by \eqref{nonmetricity_quasiplastic} such that \eqref{cond:qpd} is satisfied in order to conform to distant material parallelism, i.e., $\boldsymbol{\theta}=\boldsymbol{\zeta}=\boldsymbol{0}$, then there exist multiplicative decompositions \eqref{plastmulti} and \eqref{qpd} of the total deformation gradient into an elastic and a plastic part, and further of the plastic part into a term which relates to dislocations and other to metric anomalies. 
\end{prop}

\begin{rem}
({\it Linearization of quasi-plastic deformation.}) {\rm Consider the linearizations
\begin{subequations}
\begin{align}
 \boldsymbol{\mathcal{Q}} &\approx \boldsymbol{I}+\hat{\boldsymbol{q}}+\boldsymbol{w}\,\,\textrm{and}\\
 \boldsymbol{g} &\approx \boldsymbol{I}+2\boldsymbol{\epsilon},
 \end{align}
\end{subequations}
where $\hat{\boldsymbol{q}} \in Sym$, $\boldsymbol{w} \in Skw$ and $\boldsymbol{\epsilon} \in Sym$ such that they are all of the same order. Also assume the quasi-plastic strain ${\boldsymbol{q}}$ to be infinitesimally small of the order of $\hat{\boldsymbol{q}}$. Substituting the above approximations into \eqref{sqbq}, and collecting the leading order terms, we can identify $\hat{\boldsymbol{q}}$ with ${\boldsymbol{q}}$. The tensor $\boldsymbol{w}$ is left undetermined. The two frameworks therefore coincide in a linearized formulation.
} \label{linearqp}
\end{rem}

\begin{rem}
({\it Anisotropic distribution of point defects.}) {\rm In the introduction, we referred to certain clusters of point defects in crystals which form exotic shapes in their stable equilibrium configurations (Figures \ref{split} and \ref{tetra-penta}). A continuous distribution of such anisotropic point defects can be represented in terms of quasi-plastic deformation tensor $\boldsymbol{\mathcal{Q}}$ for a suitable symmetry class. The symmetry class of $\boldsymbol{\mathcal{Q}}$ corresponds to the structural symmetries of the shape of the point defect clusters distributed throughout the body; it is the point symmetry group of each clusters in $\mathbb{E}^3$. For example,
for the case of split-interstitials (as shown in Figure \ref{split}), each cluster is transversely isotropic with axis along the dumbbell. We can read off the transversely isotropic representation form for $\boldsymbol{\mathcal{Q}}$ from the table provided in Section 4 of Lokhin and Sedov \cite{lokhin}. This table contains forms of various invariant tensors for all the crystal symmetry classes. If the transverse isotropy axis field is given by $\boldsymbol{k}(\boldsymbol{X})$, then $\boldsymbol{Q}$ has the representation $\boldsymbol{\mathcal{Q}}=A (\boldsymbol{X}) \,\boldsymbol{G}_i\otimes\boldsymbol{G}^i + B (\boldsymbol{X}) \,\boldsymbol{k}\otimes\boldsymbol{k}$
in terms of the two scalar fields $A (\boldsymbol{X})$ and $B (\boldsymbol{X})$ and the unit vector field $\boldsymbol{k}(\boldsymbol{X})$. Note that the scalar field $B(\boldsymbol{X})$ captures the anisotropic part of $\boldsymbol{\mathcal{Q}}$. 
 }
\end{rem}


\begin{rem}
 ({\it Anisotropic thermal deformation.}) {\rm Thermal deformation can be modelled as metric anomalies in the material manifold \cite{kroner81a}. For modelling thermal deformation, the appropriate form of quasi-plastic deformation is $\boldsymbol{\mathcal{Q}}=\boldsymbol{\Lambda}\Delta T$, where $\boldsymbol{\Lambda}$ is the symmetric tensorial coefficient of thermal expansion and $\Delta T$ is the temperature change. Anisotropic deformation is characterized by appropriate forms of $\boldsymbol{\Lambda}$ chosen from the table provided in Lokhin and Sedov \cite{lokhin} for the specific symmetry class under consideration.
 }
\end{rem}

 
\section{Stress-free distribution of metric anomalies}\label{sec4}

By alternating various indices in the definition \eqref{def:nonmetricity} of the non-metricity tensor, the coefficients of the material connection can be written as \cite{schouten}
\begin{equation}
 L^p_{ij}=\Gamma^p_{ij}+W_{ij}{}^p,
\end{equation}
where
\begin{subequations}
\begin{align}
 \Gamma^p_{ij} &:= \frac{1}{2}g^{pn}(g_{ni,j}+g_{nj,i}-g_{ij,n}),\\
W_{ij}{}^p &:=  C_{ij}{}^p+ M_{ij}{}^p,\\
C_{ij}{}^p &:= g^{pk}\big(-T_{ikj}+T_{kji}-T_{jik}),\\
M_{ij}{}^p &:= \frac{1}{2} g^{pk} \big(Q_{ikj}-Q_{kji}+Q_{jik}\big) ~~~\text{and}
\end{align}
\label{def}%
\end{subequations}
$T_{ijp}    := T_{ij}{}^{k} g_{kp}$.
The functions $\Gamma^p_{ij}$ are the coefficients of the Levi-Civita connection of the metric $g_{ij}$, whereas the functions $C_{ij}{}^p$ form the components of the contortion tensor \cite{noll}.
 It is straightforward to derive the following identity relating the purely covariant components $R_{ijpl}$ and $K_{ijpl}$ of the Riemann-Christoffel curvatures of the material connection $L^{p}_{ij}$ and the Levi-Civita connection $\Gamma^p_{ij}$, respectively (cf. \cite[Part III, Section 4]{schouten}):
\begin{equation}
R_{ijpl}=K_{ijpl} +2\nabla_{[i}W_{j]pl}+ 2W_{[j|p|}{}^m\,Q_{i]m l} -2W_{[i|m l|}\, W_{j]p}{}^m + 2T_{ij}{}^m W_{mpl}
\label{identity2}
\end{equation}
or equivalently
\begin{equation}
R_{ijpl}=K_{ijpl} +2\partial_{[i}W_{j]pl} +2W_{[i|m l|}\, W_{j]p}{}^m,
\label{identity1}
\end{equation}
where $W_{ijp}:=W_{ij}{}^k g_{kp}$; $\partial$ denotes covariant differentiation with respect to the Levi-Civita connection $\Gamma^p_{ij}$. The skew part of \eqref{identity1} (or \eqref{identity2}) with respect to indices $pl$ yields the third Bianchi-Padova identity \eqref{diff:3} (recall that $K_{ij[pl]} = 0$ by definition). The symmetric part, on the other hand, provides a system of non-linear PDEs for the material metric $g_{ij}$, given various material inhomogeneity measures in terms of material curvature $R_{ijp}{}^q$, material torsion $T_{ij}{}^p$ and material non-metricity $Q_{kij}$. In fact, in the absence of inhomogeneities, \eqref{identity1} is reduced to $K_{ijpl} = 0$ which, provided that $\mathcal{B}$ is simply connected, yields the classical fundamental theorem of Riemannian geometry, i.e., there exists a sufficiently smooth diffeomorphism $\boldsymbol{\chi}:\mathcal{B}\to\mathbb{R}^3$ such that $\boldsymbol{g}=(\mbox{Grad}\boldsymbol{\chi})^T \mbox{Grad}\boldsymbol{\chi}$. Here $\mbox{Grad}$ denotes the covariant differentiation with respect to the Levi-Civita connection of the Euclidean metric $G_{ij}$.

In order to formulate the boundary value problem for the internal stress field in the body due to a distribution of material inhomogeneities, the above discussion leads us to the important interpretation of the material metric $\boldsymbol{g}$ in terms of the {\it elastic strain tensor} $\boldsymbol{E}:=\frac{1}{2}(\boldsymbol{g}-\boldsymbol{I})$. Consequently, the curvature $K_{ijpl}$ can be identified with the {\it incompatibility} of the elastic strain field $E_{ij}$. Vanishing of $K_{ijpl}$ in the absence of material inhomogeneities in a simply connected body results in the existence of an {\it elastic deformation field} $\boldsymbol{\chi}$ such that $\boldsymbol{E}=\frac{1}{2} \big((\mbox{Grad}\boldsymbol{\chi})^T \mbox{Grad}\boldsymbol{\chi}-\boldsymbol{I}\big)$ \cite{kroner81a, dewit81}.\footnote{Recall, from our discussion of quasi-plastic deformation in the last section, the tensor $\boldsymbol{H}$ which is now identified with $\mbox{Grad}\boldsymbol{\chi}$ in the absence of material inhomogeneities.} Hence, the symmetric part of \eqref{identity1} (or \eqref{identity2}) with respect to indices $pl$, along with the equilibrium equations which arise from the classical minimization problem of the functional $\int_{\mathcal{B}}W(\boldsymbol{I}+2\boldsymbol{E})dV$ over a suitable solution class of functions $\boldsymbol{E}$ (equivalently $\boldsymbol{g}$) compatible with the boundary data, constitute the boundary value problem for internal stress field generated by a given distribution of material inhomogeneities \cite{kroner81a}.

Under the assumption of absolute distant material parallelism, i.e., $R_{ijpl}=0$, and zero dislocation density, i.e., $T_{ij}{}^p=0$, \eqref{identity1} yields
\begin{equation}
0=K_{ijpl} +2\partial_{[i}M_{j]pl} +2M_{[i|m l|}\, M_{j]p}{}^m.
\label{identity4}
\end{equation}
In absence of metric anomalies, i.e., $M_{ijk}=0$, $K_{ijpl}=0$, which would imply a vanishing internal stress field if there are no external sources of stress. The contrary however is not true. We can have a non-trivial distribution of metric anomalies which lead to $K_{ijpl}=0$. In rest of this section our aim is to obtain the general form of the non-metricity tensor $Q_{kij}$, which when substituted into \eqref{identity4} gives $K_{ijpl}=0$, under the assumption that both elastic strain $E_{ij}$ and $Q_{kij}$ are small and are of the same order. Under such an assumption, with $K_{ijpl}=0$, \eqref{identity4} can be linearized to obtain 
\begin{equation}
 \partial_i M_{jpl}-\partial_j M_{ipl}=0
 \label{identity3}
 \end{equation}
with $M_{j(pl)} = 2Q_{jpl}$ and $M_{j[pl]} = Q_{[pl]j}$. Taking the symmetric part of \eqref{identity3} with respect to $pl$, we obtain
\begin{equation}
 \partial_i Q_{jpl}-\partial_j Q_{ipl}=0.
\end{equation}
The torsion and curvature corresponding to both the connections $L^p_{ij}$ and $\Gamma^p_{ij}$ are identically zero. The last equation then implies, from Poincar\'e lemma, assuming $\mathcal{B}$ to be simply connected, that there exists a symmetric matrix field $[S_{pl}]$ over $\mathcal{B}$ such that
\begin{equation}
 Q_{ipl}=\partial_i S_{pl}.
 \label{rep}
\end{equation}
On the other hand, the skew part of \eqref{identity3} with respect to $pl$, i.e., $\partial_iQ_{[pl]j}-\partial_jQ_{[pl]i}=0$, after substituting from \eqref{rep}, yields
\begin{equation}
 \partial_i\partial_p S_{lj}- \partial_j\partial_pS_{li} - \partial_i\partial_lS_{pj} +\partial_j\partial_lS_{pi}=0.
\end{equation}
This relation implies that there exists a sufficiently smooth vector field $A_l$ over $\mathcal{B}$, such that
\begin{equation}
 S_{pl}=\partial_{(p}A_{l)}.
\end{equation}
Hence,
\begin{equation}
 Q_{ipl}=\partial_i\partial_{(p}A_{l)}
\end{equation}
is the most general form of stress-free distribution of metric anomalies under the assumptions made above. 
We summarize the result as
\begin{prop}\label{nilnm}
In absence of dislocations and disclinations in a simply connected material body $\mathcal{B}$, if the elastic strain and non-metricity tensor are assumed to be small and of the same order, then there exists a sufficiently smooth vector field $A_j$ over $\mathcal{B}$ such that the stress-free distribution of metric anomalies is given by $Q_{kij}=\partial_k\partial_{(i}A_{j)}$.
\end{prop}

\section{Concluding remarks}\label{sec5}
Metric anomalies in a materially uniform simple elastic solid, in general, give rise to a distribution of metrical disclinations which requires the inner product of tangent vectors in the material space to be path dependent. For an unambiguous definition of crystallinity at all points in a crystalline elastic solid absolute distant parallelism must be maintained and curvature anomalies (disclinations) should  disappear. With this geometrical viewpoint, metric anomalies in a materially uniform elastic crystalline solid have to be irrotational. Weyl geometry furnishes an isotropic non-metricity field under the assumption of irrotationality and is hence insufficient to represent anisotropic metric anomalies. The quasi-plastic strain formulation is sufficiently general but provides no basis for a multiplicative decomposition of the deformation gradient. The quasi-plastic deformation framework, as introduced in this article, overcomes this shortcoming but only at the cost of satisfying additional equations. It allows for a multiplicative decomposition of the material uniformity field (plastic deformation tensor) into a deformation tensor, due to dislocation distribution, and the quasi-plastic deformation tensor which represents the non-metricity in the material space. The present work can be used to understand the geometrical nature of a wide variety of anisotropic metric anomalies as they may appear in a distribution of point defects, anisotropic thermoelasticity, anisotropic biological growth, etc. We defer detailed applications of our framework to a future work.


\bibliographystyle{plain}
\bibliography{inhomogeneity}

\end{document}